\documentclass[aos,preprint]{imsart}

\usepackage[normalem]{ulem}
\usepackage{soul}

\RequirePackage[OT1]{fontenc}
\RequirePackage{amsthm,amsmath,bm}
\RequirePackage[numbers]{natbib}
\RequirePackage[colorlinks,citecolor=blue,urlcolor=blue]{hyperref}
\usepackage[psamsfonts]{amssymb}
\usepackage{graphicx}
\usepackage{epstopdf}
\usepackage{epsfig}
\usepackage{subfigure}
\usepackage{natbib}
\usepackage{threeparttable}

\newcommand{\bDelta}{{\bm\Delta}}
\newcommand{\bGamma}{{\bm\Gamma}}

\newcommand{\bSigma}{{\bm\Sigma}}

\def\d{{\bm d}}

\def\v{{\bm v}}

\def\x{{\bm x}}
\def\z{{\bm z}}
\def\0{{\bm 0}}

\def\A{{\bm A}}
\def\B{{\bm B}}
\def\C{{\bm C}}
\def\D{{\bm D}}
\def\E{{\bm E}}
\def\H{{\bm H}}
\def\S{{\bm S}}
\def\I{{\bm I}}
\def\M{{\bm M}}
\def\Q{{\bm Q}}

\def\X{{\bm X}}
\def\U{{\bm U}}
\def\V{{\bm V}}
\def\W{{\bm W}}
\def\X{{\bm X}}
\def\Y{{\bm Y}}
\def\Z {{\bm Z}}

\def\tr{\mbox{tr}}
\def\diag{\mbox{diag}}
\startlocaldefs
\numberwithin{equation}{section}
\theoremstyle{plain}
\newtheorem{theorem}{Theorem}[section]
\newtheorem{lemma}{Lemma} 
\newtheorem{remark}{Remark}  
\newtheorem{corollary}{Corollary}[section]
\endlocaldefs

\begin{document}
\begin{frontmatter}
\title{Detection of the number of principal components by extended AIC-type method}
\runtitle{Extended AIC-type method}

\begin{aug}
\author{\fnms{Jianwei} \snm{Hu}\thanksref{t1}\ead[label=e1]{jwhu@mail.ccnu.edu.cn}},
\author{\fnms{Jingfei} \snm{Zhang}\thanksref{t2}\ead[label=e2]{ezhang@bus.miami.edu}}
\and
\author{\fnms{Ji} \snm{Zhu}\thanksref{t3}\ead[label=e3]{jizhu@umich.edu}}
\runauthor{Hu, Zhang and Zhu}
\affiliation{Central China Normal University\thanksmark{t1}, University of Miami\thanksmark{t2}\\ and University of Michigan\thanksmark{t3}}
\address{Department of Statistics\\
Central China Normal University\\
Wuhan, 430079\\
China\\
\printead{e1}
}

\address{Department of Management Science\\
University of Miami\\
Coral Gables, FL 33124\\
USA\\
\printead{e2}
}

\address{Department of statistics\\
University of Michigan\\
Ann Arbor, MI 48109\\
USA\\
\printead{e3}}

\end{aug}
\begin{abstract}
Estimating the number of principal components is one of the fundamental problems in many scientific fields such as signal processing (or the spiked covariance model).
In this paper, we first demonstrate that, for fixed $p$, any penalty term of the form $k'(p-k'/2+1/2)C_n$ may lead to an asymptotically consistent estimator under the condition that $C_n\to\infty$ and $C_n/n\to0$. Compared with the condition in \citep{Zhao:Krishnaiah:Bai:1986}, i.e., $C_n/\log\log n\to\infty$ and $C_n/n\to0$, this condition is significantly weakened. Then we propose to select the number of signals $k$ by the iterated logarithm penalty (ILP).
We also extend our results to the case $n,p\to\infty$, with $p/n\to c>0$. In this case, for increasing $k$, we first investigate the limiting laws for the leading eigenvalues of the sample covariance matrix $\S_n$ under the condition that $\lambda_k>1+\sqrt{c}$, which extend the results in \citep{Bai:Choi:Fujikoshi:2018} and \citep{Cai:Han:Pan:2019}. This includes the case $\lambda_k\rightarrow\infty$. At low SNR, since the AIC tends to underestimate the number of signals $k$,  the AIC should be re-defined in this case. As a natural extension of the AIC for fixed $p$, we propose the extended AIC (EAIC), i.e., the AIC-type method with tuning parameter $\gamma=\varphi(c)=1/2+\sqrt{1/c}-\log(1+\sqrt{c})/c$, and demonstrate that the EAIC-type method, i.e., the AIC-type method with tuning parameter $\gamma>\varphi(c)$, can select the number of signals $k$ consistently. As a result, in the following two cases, (1) $p$ fixed, $n\to\infty$,
(2) $n,p\to\infty$ with $p/n\to 0$, if the AIC is defined as the degeneration of the EAIC in the case $n,p\to\infty$ with $p/n\to c>0$, i.e., $\gamma=\lim_{c\rightarrow 0+}\varphi(c)=1$, then we have essentially demonstrated that, the AIC tends to overestimate $k$. To achieve the consistency of the AIC-type method in the above two cases, $\gamma>1$ is required. On the other hand, in the case $n,p\to\infty$, with $p/n\to c>0$, we have $\varphi(c)<1$. Then in this case, we have actually explained why the AIC tends to underestimate $k$. Moreover, we show that the EAIC-type method is essentially tuning-free and outperforms the well-known KN estimator proposed in \citep{Kritchman:Nadler:2008} and the BCF estimator proposed in \citep{Bai:Choi:Fujikoshi:2018}. Numerical studies indicate that the proposed method works well.\\
\end{abstract}
\begin{keyword}[class=MSC]
\kwd[Primary ]{62E20}
\kwd[; secondary ]{60G05}
\end{keyword}

\begin{keyword}
\kwd{Consistency}
\kwd{minimum description length}
\kwd{signal-to-noise ratio}
\kwd{spiked covariance model}
\kwd{Tracy-Widom distribution}
\end{keyword}

\end{frontmatter}

\section{Introduction}
\label{section:introduction}

Detection of the number of principal components from a noisy data is one of the fundamental problems in many scientific fields. It is often the starting point for the signal parameter estimation problem such as signal processing \citep{Wax:Kailath:1985}, wireless communications \citep{Nicoli:Simeone:Spagnolini:2003}, array processing \citep{Bohme:1991} and finance \citep{Bai:Ng:2002} to list a few and has primary importance \citep{Anderson:2003}. The most common approach to solving this problem is by using information theoretic criteria, and in particular minimum description length (MDL) \citep{Rissanen:1978}, Bayesian information criterion (BIC) and Akaike information criterion (AIC) \citep{Wax:Kailath:1985}.

Consider the model
\begin{eqnarray}
\label{eq:signal}
\x(t)=\M \mathbf{s}(t)+\mathbf{n}(t)=(\x_1^\top(t),\x_2^\top(t))^\top
\end{eqnarray}
where $k<p$, $\M=(\M(\Psi_1),\cdots,\M(\Psi_k))$, $\mathbf{s}(t)=(\mathbf{s}_1(t),\cdots,\mathbf{s}_k(t))^\top$ and $\mathbf{n}(t)=(\mathbf{n}_1(t),\cdots,\mathbf{n}_p(t))^\top$. In (\ref{eq:signal}), $\mathbf{n}(t)$ is the noise vector distributed independent of $\mathbf{s}(t)$ as multivariate normal with mean vector $0$ and covariance matrix $\sigma^2 \I_p$. $\mathbf{s}(t)$ is distributed as multivariate normal with mean $0$ and nonsingular matrix $\bm\Omega$ and $\M(\bm\Psi_i): p\times 1$ is a vector of functions of the elements of unknown vector $\bm\Psi_i$ associated with $i$-th signal. Then, the covariance matrix $\bm\Sigma$ of $\x(t)$ is given by
\begin{eqnarray}
\label{eq:spike}
\bm\Sigma=\M\bm\Omega \M^\top+\sigma^2 \I_p=\left(\begin{array}{cc}
                              \bSigma_{11} & \bSigma_{12} \\
                              \bSigma_{21} & \bSigma_{22} \\
                            \end{array}\right).
\end{eqnarray}
We assume that $\x(t_1),\cdots, \x(t_n)$ are independent observations on $\x(t)$. Let $\lambda_1\geq\cdots\geq\lambda_k>\lambda_{k+1}=\cdots=\lambda_{p}=\lambda=\sigma^2$ denote the eigenvalues of $\bSigma$. Note that (\ref{eq:spike}) is also called the spiked covariance model in \cite{Johnstone:2001}.

Using the well-known spectral representation theorem from linear algebra, we can express $\Sigma$ as
\begin{equation}
\label{eq:gamma}
\bSigma=\sum_{i=1}^k(\lambda_i-\lambda)\bGamma_i\bGamma_i^\top+\lambda \I_p,
\end{equation}
where $\bGamma_i$ is the eigenvector of $\bSigma$ with eigenvalue $\lambda_i$. Denoting by $\bm\theta$ the parameter vector of the model, it follows that
\[
\bm\theta^\top=(\bm\theta_{1}^\top,\bm\theta_{2}^\top),
\]
where $\bm\theta_{1}^\top=(\lambda_1,\cdots,\lambda_k,\lambda)$ and $\bm\theta_{2}^\top=(\bGamma_1^\top,\cdots,\bGamma_k^\top)$.

With this  parameterization we now proceed to the derivation of the information theoretic criteria for the detection problem. Since the observation are regarded as statistically independent Gaussian random vectors with zero mean, their joint probability density is given by
\[
f(\X\mid\theta)=\prod_{i=1}^nf(\x(t_i)\mid\bm\theta)=\prod_{i=1}^n\frac{1}{(2\pi)^{p/2}\mid\bSigma\mid^{1/2}}\exp{-\frac{1}{2}\x^\top(t_i)\bSigma^{-1}\x(t_i)},
\]
where $\X^\top=(\x^\top(t_1),\cdots,\x^\top(t_n))$.

Taking the logarithm, the log-likelihood function is given by
\begin{eqnarray*}
\log f(\X\mid\theta)&=&\sum_{i=1}^n\log f(\x(t_i)\mid\bm\theta)\\
&=&\log \prod_{i=1}^n\frac{1}{(2\pi)^{p/2}\mid\bSigma\mid^{1/2}}\exp{-\frac{1}{2}\x^\top(t_i)\bSigma^{-1}\x(t_i)}\\
&=&-\frac{1}{2}n\log \mid\bSigma\mid-\frac{1}{2}n\tr(\bSigma^{-1}\S_n)-\frac{1}{2}np\log(2\pi).
\end{eqnarray*}
where $\S_n=\frac{1}{n}\sum_{i=1}^n\x(t_i)\x(t_i)^\top=\frac{1}{n}\X^\top\X$ is the sample covariance matrix.

Let $\bDelta=\diag\{\lambda_1,\cdots,\lambda_p\}$ and $\bSigma=\bGamma\bDelta\bGamma^\top$, where $\bGamma$ is orthogonal and its column $\bGamma_i$ is the eigenvector of $\bSigma$  with eigenvalue $\lambda_i$. Also,  denote by $d_1\geq\cdots\geq d_p$ the eigenvalues of $\S_n$. Let $\D=\diag\{d_1,\cdots,d_p\}$ and
\begin{equation}
\label{eq:C}
\S_n=\C\D\C^\top\triangleq\left(\begin{array}{cc}
                              \S_{11} & \S_{12} \\
                              \S_{21} & \S_{22} \\
                            \end{array}\right)=\frac{1}{n}\left(\begin{array}{cc}
                              \X_1^\top\X_1 & \X_1^\top\X_2 \\
                              \X_2^\top\X_1 & \X_2^\top\X_2 \\
                            \end{array}\right),
\end{equation}
where $\C$ is orthogonal and its column $\C_j$ is the eigenvector of $\S_n$  with eigenvalues $d_j$.

In the signal processing literature, the AIC estimates $k$ by maximizing $\ell(k')$:
\[
\hat{k}=\arg\max_{k'}\ell(k'),
\]
where
\[
\ell(k')=-\frac{1}{2}n(\sum_{i=1}^{k'}\log d_i+(p-k')\log\hat{\lambda}_{k'})-k'(p-k'/2+1/2).
\]
Here $\hat{\lambda}_{k'}=1/(p-k')\sum_{i=k'+1}^pd_i$.

The BIC (or the MDL) estimates $k$ by maximizing $\ell(k')$:
\[
\hat{k}=\arg\max_{k'}\ell^B(k'),
\]
where
\[
\ell^B(k')=-\frac{1}{2}n(\sum_{i=1}^{k'}\log d_i+(p-k')\log\hat{\lambda}_{k'})-\frac{1}{2}k'(p-k'/2+1/2)\log n.
\]

Note that the AIC and the BIC are slightly different from that of \cite{Wax:Kailath:1985}, where the number of unknown parameters was $k'(p-k'/2)$ and was corrected by \cite{Zhao:Krishnaiah:Bai:1987}.

The first consistency proof of the BIC was established in \cite{Wax:Kailath:1985}. By using  Wilks' theorem, \citeauthor{Wax:Kailath:1985} \cite{Wax:Kailath:1985} showed that any penalty term of the form $k'(p-k'/2)C_n$ may lead to an asymptotically consistent estimator under the condition that $C_n\to\infty$ and $C_n/n\to0$. Later, it was demonstrated that Wilks' theorem cannot be used in this problem in \cite{Zhao:Krishnaiah:Bai:1987}. By using Taylor's expansion,  \citeauthor{Zhao:Krishnaiah:Bai:1986} \cite{Zhao:Krishnaiah:Bai:1986} gave an alternative proof and showed that, any penalty term of the form $k'(p-k'/2+1/2)C_n$ may lead to an asymptotically consistent estimator under the condition that $C_n/\log\log n\to\infty$ and $C_n/n\to0$. Although \citeauthor{Zhao:Krishnaiah:Bai:1986} \cite{Zhao:Krishnaiah:Bai:1986} proved that the BIC is consistent, the BIC fails to detect signals at low signal-to-noise ratio (SNR), hence underestimating the number of signals at small sample size. In contrast, while the AIC is able to detect low SNR signals, it has a non-negligible probability to overestimate the number of signals and thus is not consistent. To remedy the shortcoming of the AIC, \citeauthor{Nadler:2010} \cite{Nadler:2010} proposed an modified AIC. The only difference between these two methods lies in that the penalty of the modified AIC is two times that of the AIC. In the presence of pure noise with no signals (i.e., $k=0$), \citeauthor{Nadler:2010} \cite{Nadler:2010} showed that the modified AIC had a negligible overestimation probability for large $n$. However, for $k>0$, \citeauthor{Nadler:2010} \cite{Nadler:2010} did not give an explanation why the modified AIC had a negligible overestimation probability. On the other hand, for small and medium $n$, the modified AIC tends to underestimate the number of signals $k$ in our simulations.

\citeauthor{Kritchman:Nadler:2008} \cite{Kritchman:Nadler:2008} and \citeauthor{Kritchman:Nadler:2009} \cite{Kritchman:Nadler:2009} proposed a quite different method for estimating the number of signals, via a sequence of hypothesis tests, at each step testing the significance of the $k$-th eigenvalue as arising from a signal. The main tools used in these two papers are recent results from random matrix theory regarding both the distribution of noise eigenvalues and of signal eigenvalues in the presence of noise. In the absence of signals, the matrix $n\S_n$ follows a Wishart distribution \citep{Wishart:1928} with parameter $n, p$. In the joint limit $n,p\to\infty$, with $p/n\to c>0$, the distribution of the largest eigenvalue of $\S_n$ converges to a Tracy-Widom distribution \citep{Johansson:2000},\citep{Johnstone:2001},\cite{ElKaroui:2006},\citep{Johnstone:Lu:2009},\citep{Ma:2012}. For fixed $p$, although  \citeauthor{Kritchman:Nadler:2009}  \cite{Kritchman:Nadler:2009} proved the strong consistency of their estimator, our simulations show that when the SNR is low, for fixed $p$ and medium $n$, the KN tends to underestimate the number of signals $k$.

Before further proceeding, we mention that a similar, if not identical problem, also appears in other literatures, likelihood ratio test statistic
\citep{Muirhead:2002}, Kac-Rice test \citep{Choi:Taylor:Tibshirani:2017}. For some other work in this topic, we refer to \cite{Paul:2007}, \cite{Nadler:2008}, \cite{Birnbaum:Johnstone:Nadler:Paul:2013}, \cite{Passemier:Yao:2014}, \cite{Cai:Ma:Wu:2015}, \cite{Bao:Pan:Zhou:2015}, \cite{Bai:Choi:Fujikoshi:2018}  and so on.

In this paper, we first demonstrate that any penalty term of the form $k'(p-k'/2+1/2)C_n$ may lead to an asymptotically consistent estimator under the condition that $C_n\to\infty$ and $C_n/n\to0$. Compared with the condition in \citep{Zhao:Krishnaiah:Bai:1986}, i.e., $C_n/\log\log n\to\infty$ and $C_n/n\to0$, this condition is significantly weakened. Then we propose to select the number of signals $k$ by the iterated logarithm penalty (ILP). Moreover, we also extend our results to the case $n,p\to\infty$, with $p/n\to c>0$. In this case, for increasing $k$, we first investigate the limiting laws for the leading eigenvalues of the sample covariance matrix $\S_n$ under the condition that $\lambda_k>1+\sqrt{c}$, which extend the results in \citep{Bai:Choi:Fujikoshi:2018} and \citep{Cai:Han:Pan:2019}. This includes the case $\lambda_k\rightarrow\infty$.
At low SNR, since the AIC tends to underestimate the number of signals $k$,  the AIC should be re-defined in this case. As a natural extension of the AIC for fixed $p$, we propose the extended AIC (EAIC), i.e., the AIC-type method with tuning parameter $\gamma=\varphi(c)=1/2+\sqrt{1/c}-\log(1+\sqrt{c})/c$, and demonstrate that the EAIC-type method, i.e., the AIC-type method with tuning parameter $\gamma>\varphi(c)$, can select the number of signals $k$ consistently. Moreover, we show that the EAIC-type method is essentially tuning-free and outperforms the well-known KN estimator proposed in \cite{Kritchman:Nadler:2008} and the BCF estimator proposed in \cite{Bai:Choi:Fujikoshi:2018}.

For the remainder of the paper, we proceed as follows.
In Section \ref{section:ILP}, we demonstrate that any penalty term of the form $k'(p-k'/2+1/2)C_n$ may lead to an asymptotically consistent estimator under the condition that $C_n\to\infty$ and $C_n/n\to0$.
We extend our results to the case $n,p\to\infty$, with $p/n\to c>0$ in Section \ref{section:extension}. The numerical studies are given in Section \ref{section:experiments}.
Some further discussions are made in Section \ref{section:discussion}.
All proofs are given in Section \ref{section:proof}.

\section{Fixed p}
\label{section:ILP}
In this section, we work under the classical setting where $p$, $k$ and $\lambda_i$ are all fixed as $n \rightarrow \infty$. We establish the consistency
of \eqref{eq:penalized:ILP13} in the sense that it chooses the correct $k$ with probability
tending to one when $n$ goes to infinity.

We consider the following general criterion:
\begin{equation}
\label{eq:penalized:ILP14}
\hat{k}=\arg\max_{k'}\ell(k'),
\end{equation}
where
\begin{equation}
\label{eq:penalized:ILP21}
\ell(k')=-\frac{1}{2}n(\sum_{i=1}^{k'}\log d_i+(p-k')\log\hat{\lambda}_{k'})-k'(p-k'/2+1/2)C_n,
\end{equation}
where $C_n\to\infty$ and $C_n/n\to0$. Here $\hat{\lambda}_{k'}=1/(p-k')\sum_{i=k'+1}^pd_i$.

Throughout this paper, we assume that $\lambda$ is known. Without loss of generality, we assume that $\lambda=1$. The signal-to-noise ratio is defined as
\[
SNR=\frac{\lambda_k-\lambda}{\lambda}=\frac{\lambda_k}{\lambda}-1=\lambda_k-1.
\]

Define $\hat{\lambda}=1/(p-k)\sum_{i=k+1}^pd_i$. Note that for $i=1,2,\cdots, k$, $d_i$ and $\hat{\lambda}$ are the maximum likelihood estimates of $\lambda_i$ and $\lambda$, respectively. The following result which shows how close is the eigenvalue of $\Sigma$ to its maximum likelihood estimate.
\begin{lemma}
\label{lemma:3} For $i=1,2,\cdots, p$,
$$(\lambda_i-d_i)^2=O_p(p/n)=O_p(1/n), $$
$$ (p-k)(\lambda-\hat{\lambda})^2=O_p(p^2/n)=O_p(1/n).$$
\end{lemma}

Noting that for fixed $p$, it is difficult to demonstrate the consistency of \eqref{eq:penalized:ILP14} directly (see the proof of Theorem \ref{theorem:4}). Similar to the discussion in \cite{Zhao:Krishnaiah:Bai:1986} (see also \cite{Anderson:1963}) , we consider an alternative method.

\begin{lemma}
\label{lemma:4}
When $k'$ is equal to $k$, $\ell(k')$ defined in \eqref{eq:penalized:ILP21} can be re-written as:
\begin{equation}
\label{eq:penalized:ILP22}
\ell(k)=-\frac{1}{2}n\sum_{i=1}^k\log d_i-\frac{1}{2}n\sum_{i=k+1}^p(d_i-1)-k(p-k/2+1/2)C_n.
\end{equation}
where $C_n\to\infty$ and $C_n/n\to0$.
\end{lemma}

Then we estimate $k$ by maximizing $\tilde{\ell}(k')$:
\begin{equation}
\label{eq:penalized:ILP13}
\hat{k}=\arg\max_{k'}\tilde{\ell}(k'),
\end{equation}
where \begin{equation}
\label{eq:penalized:ILP}
\tilde{\ell}(k')=-\frac{1}{2}n\sum_{i=1}^{k'}\log d_i-\frac{1}{2}n\sum_{i=k'+1}^p(d_i-1)-k'(p-k'/2+1/2)C_n.
\end{equation}
where $C_n\to\infty$ and $C_n/n\to0$.

Next, we establish the consistency of \eqref{eq:penalized:ILP13}.

\begin{theorem}
\label{theorem:2}
Let $\tilde{\ell}(k')$ be the penalized likelihood function defined in (\ref{eq:penalized:ILP}).\\
For $k'<k$,
\[
P(\tilde{\ell}(k)>\tilde{\ell}(k'))\rightarrow1.
\]
For $k'>k$,
\[
P(\tilde{\ell}(k)>\tilde{\ell}(k'))\rightarrow1.
\]
\end{theorem}

Especially, we propose to select the number of signals $k$ by the iterated logarithm penalty (ILP). That is, we estimate $k$ by maximizing $\ell(k')$:
\begin{equation}
\label{eq:penalized:ILP12}
\hat{k}=\arg\max_{k'}\ell(k'),
\end{equation}
where
\begin{equation}
\label{eq:penalized:ILP1}
\ell(k')=-\frac{1}{2}n(\sum_{i=1}^{k'}\log d_i+(p-k')\log\hat{\lambda}_{k'})-\gamma k'(p-k'/2+1/2)\log\log n,
\end{equation}
where $\gamma>0$ is a tuning parameter.

As an alternative method, we may also estimate $k$ by maximizing $\tilde{\ell}(k')$:
\begin{equation}
\label{eq:penalized:ILP15}
\hat{k}=\arg\max_{k'}\tilde{\ell}(k'),
\end{equation}
where \begin{equation}
\label{eq:penalized:ILP16}
\tilde{\ell}(k')=-\frac{1}{2}n\sum_{i=1}^{k'}\log d_i-\frac{1}{2}n\sum_{i=k'+1}^p(d_i-1)-\gamma k'(p-k'/2+1/2)\log\log n,
\end{equation}
where $\gamma>0$ is a tuning parameter.

Simulations show that \eqref{eq:penalized:ILP12} and \eqref{eq:penalized:ILP15} behave almost the same. For simplicity and comparisons with other methods, we consider \eqref{eq:penalized:ILP12} in our simulations. For simplicity, we may use $\gamma=1$ in our simulations which gives good performance.

\begin{remark} \citeauthor {Zhao:Krishnaiah:Bai:1986}  \cite{Zhao:Krishnaiah:Bai:1986} demonstrated that any penalty term of the form $k'(p-k'/2+1/2)C_n$ leads to an asymptotically consistent estimator under the condition that  $C_n/\log\log n\to\infty$ and $C_n/n\to0$. Theorem \ref{theorem:2} shows that this condition can be significantly weakened as $C_n\to\infty$ and $C_n/n\to0$. To show that our results are correct, we also consider different $C_n$'s in our simulations in Section \ref{section:experiments}, such as $ILP^{\frac{1}{2}}$, which estimates $k$ by maximizing $\tilde{\ell}(k')$:
\[
\hat{k}=\arg\max_{k'}\tilde{\ell}(k'),
\]
where \[
\tilde{\ell}(k')=-\frac{1}{2}n\sum_{i=1}^{k'}\log d_i-\frac{1}{2}n\sum_{i=k'+1}^p(d_i-1)-\gamma k'(p-k'/2+1/2)(\log\log n)^{\frac{1}{2}}.
\]
\end {remark}

\section{Infinite p}
\label{section:extension}
In this section, we allow $p$, $k$ and $\lambda_k$ grow as $n$ grows.

\subsection{Asymptotic behavior of the spiked sample eigenvalues}

In this subsection, in the case $n,p\to\infty$, with $p/n\to c>0$, for increasing $k$, we investigate the limiting laws for the leading eigenvalues of the sample covariance matrix $\S_n$ under the condition that $\lambda_k>1+\sqrt{c}$, which extend the results in \citep{Bai:Choi:Fujikoshi:2018} and \citep{Cai:Han:Pan:2019}. This includes the case $\lambda_k\rightarrow\infty$.

We wish to point out that the empirical spectral distribution (ESD) $F(x)$ of $\bSigma$ converges weakly to the limiting spectral distribution (LSD) $H(x)$ as $p\rightarrow\infty$, that is,
\[
\sup_x \mid F(x)-H(x)\mid=\frac{k}{p},
\]
where $F(x)=\frac{1}{p}\sum_{i=1}^pI_{(-\infty,x]}(\lambda_i)$, $H(x)=I(1\leq x)$ and $I_A(\cdot)$ is the indicator function of $A$.

Note that the ESD of $\S_n$ is given by
\[
F_n(x)=\frac{1}{p}\sum_{i=1}^pI_{(-\infty,x]}(d_i).
\]
With probability $1$,
\[
F_n(x)\stackrel{d}{\rightarrow}F_c(x).
\]
Here, for $0<c\leq 1$, $F_c(x)$ is given as
\[
F_c'(x)=f_c(x)=\left\{
                      \begin{array}{ll}                 \frac{1}{2\pi xc}\sqrt{(b-x)(x-a)}, & \rm{if}\,\ x\in (a,b), \\
                        0, & \rm{otherwise},
                     \end{array}
                   \right.
\]
where $a=(1-\sqrt{c})^2$ and $b=(1+\sqrt{c})^2$.

If $c>1$, $F_c(x)$ has a point mass $1-1/c$ at the origin, that is,

\[
F_c(x)=\left\{
                      \begin{array}{ll}                 0, & \rm{if}\,\, x<0, \\
                        1-1/c, & \rm{if}\,\, 0\leq x<a,\\
                        1-1/c+\int_a^xf_c(t)dt, & \rm{if}\,\, a\leq x<b,
                     \end{array}
                   \right.
\]
where $a$ and $b$ are the same as in the case $0<c\leq 1$. We remark that $\int_a^bf_c(t)dt=1$ or $1/c$ according to $0<c\leq 1$ or $c>1$ respectively.

By a result of \citeauthor{Silverstein:1995} \cite {Silverstein:1995} (see also \citeauthor{Bai:Choi:Fujikoshi:2018} \cite{Bai:Choi:Fujikoshi:2018}),
\[
\sup_x \mid F_n(x)-F_c(x)\mid=O_p(\frac{k}{n}).
\]
This implies that the ESD of $\S_n$ also converges weakly to $F_c(x)$ as $n\rightarrow\infty$ under the condition where the number of spikes, $k$, is fixed or $k=o(n)$. From the MP law, we have the easy consequence that if $i/p\rightarrow\alpha\in (0,1)$, then $d_i\stackrel{a.s.}{\rightarrow}\mu_{1-\alpha}$, where $\mu_{\alpha}$ is the $\alpha$-th quantile of the MP law, that is $F_c(\mu_{\alpha})=\alpha$.

The $i$-th largest eigenvalue, $\lambda_i$, is said to be a distant spiked eigenvalue if $\psi'(\lambda_i)>0$ where $\psi(\lambda_i)=\lambda_i+\frac{c\lambda_i}{\lambda_i-1}$. Equivalently, $\lambda_i$ is a distant spiked eigenvalue if $\lambda_i>1+\sqrt{c}$.

Define $x(t)$ as
\[ \x(t)=\bSigma^{1/2}\z(t),\]
where $\z(t)$ is distributed with mean $0$ and covariance matrix $\I_p$. Thus, the covariance matrix of $\x(t)$ is given by $\bSigma$.

For fixed $k$, under the assumption that $\lambda_1$ is bounded,  according to \cite{Bai:Choi:Fujikoshi:2018}, the following result holds.
\begin{lemma}
\label{lemma:bai:fujikoshi:choi1}
In the case $n,p\to\infty$ with $p/n\to c>0$, suppose that for any $1\leq i\leq n$, $1\leq j\leq p$, $E((\z(t_i))_j^4)\leq C$, $\lambda_1$ is bounded and $\lambda=1$.
If $\lambda_i$ is a distant spiked eigenvalue, we have
\[
d_i\stackrel{a.s.}{\longrightarrow} \psi(\lambda_i).
\]
\end{lemma}

The above definition and result are a special case of a more general definition and result in \cite{Bai:Yao:2012} and \cite{Baik:Silverstein:2006}.

For fixed $k$, under the assumption that $\lambda_k\rightarrow\infty$ as $p\rightarrow\infty$, a new limiting result for the distant spiked eigenvalues was also established in \cite{Bai:Choi:Fujikoshi:2018}.

\begin{lemma}
\label{lemma:bai:fujikoshi:choi2}
In the case $n,p\to\infty$ with $p/n\to c>0$, suppose that for any $1\leq i\leq n$, $1\leq j\leq p$, $E((\z(t_i))_j^4)\leq C$ and $\lambda=1$. If $\lambda_k\rightarrow\infty$ as $p\rightarrow\infty$, for  any $1\leq i\leq k$, we have
\[
\frac{d_i}{\psi(\lambda_i)}\stackrel{a.s.}{\longrightarrow}1.
\]
\end{lemma}

\if

For fixed $k$, from Lemmas \ref{lemma:bai:fujikoshi:choi1} and \ref{lemma:bai:fujikoshi:choi2}, under the assumption that
\begin{enumerate}
\item[(1)] at least one $\lambda_i$ ($1\leq i\leq k-1$) satisfying $\lambda_i\rightarrow\infty$ as $p\rightarrow\infty$,
\item[(2)]$1+\sqrt{c}<\lambda_k<\infty$,
\end{enumerate}
for any $1\leq i\leq k$,
\[
\frac{d_i}{\psi(\lambda_i)}\stackrel{a.s.}{\longrightarrow}1.
\]
is still not established.

\fi

For increasing $k$, under the assumption that $\lambda_k\rightarrow\infty$ as $p\rightarrow\infty$, \citeauthor{Cai:Han:Pan:2019} \cite{Cai:Han:Pan:2019} established the following result.
\begin{lemma}
\label{lemma1:cai:han:pan}
In the case $n,p\to\infty$ with $p/n\to c>0$, suppose that for any $1\leq i\leq n$, $1\leq j\leq p$, $E((\z(t_i))_j^4)\leq C$ and $\lambda_{k+1}\geq\cdots\geq\lambda_{p}$ are bounded. If $\lambda_k\rightarrow\infty$ as $p\rightarrow\infty$ and $k=\min\{o(n^{1/6}), o(\lambda_i^{-\frac{1}{2}})\}$, for  any $1\leq i\leq k$, we have
\[
\frac{d_i}{\lambda_i}-1=\max\{O_p(\frac{1}{\lambda_i}), O_p(\frac{k^4}{n})\}.
\]
\end{lemma}

Based on Lemma \ref{lemma1:cai:han:pan}, \citeauthor{Cai:Han:Pan:2019} \cite{Cai:Han:Pan:2019} also established the central limit theorem for increasing $k$, under the assumption that $\lambda_k\rightarrow\infty$ as $p\rightarrow\infty$.

\begin{lemma}
\label{lemma2:cai:han:pan}
In the case $n,p\to\infty$ with $p/n\to c>0$, suppose that for any $1\leq i\leq n$, $1\leq j\leq p$, $E((\z(t_i))_j^4)\leq C$ and $\lambda_{k+1}\geq\cdots\geq\lambda_{p}$ are bounded, there exists a positive constant $C$ not depending on $n$ such that $\frac{\lambda_{i-1}}{\lambda_i}\geq C>1$, $i=1,2,\dots,k$. If $\lambda_k\rightarrow\infty$ as $p\rightarrow\infty$ and $k=\min\{o(n^{1/6}), o(\lambda_i^{-\frac{1}{2}})\}$, for  any $1\leq i\leq k$, we have
\[
\frac{\sqrt{n}(d_i-\psi(\lambda_i))}{\psi(\lambda_i)}\stackrel{d}{\longrightarrow} N(0, \sigma_i^2),
\]
where $\sigma_i^2$ is defined in \cite{Cai:Han:Pan:2019}.
\end{lemma}

For more work on this topic, we refer to \cite{Jonsson:1982}, \cite{Johansson:1998},  \cite{Bai:Silverstein:2004}, \cite{Nadakuditi:Edelman:2008}, \cite{Onatski:Moreira:Hallin:2013}, \cite{Wang:Yao:2013}, \cite{Wang:Silverstein:Yao:2014}, \cite{Passemier:Mckay:Chen:2015}, \cite{Wang:Fan:2017} and \cite{Li:Han:Yao:2019}.

For increasing $k$, under the assumption that $\lambda_k>1+\sqrt{c}$, we will extend the results in \citep{Bai:Choi:Fujikoshi:2018} and \citep{Cai:Han:Pan:2019}.

Our first result gives the limits for the spiked eigenvalues of $\S_n$, $d_1\geq\cdots\geq d_k$, which will be used to establish the consistency of the following EAIC-type method.
\begin{theorem}\label{theorem:a.s.consistency}
Suppose that $k=o(n^{\frac{1}{3}})$ and $\frac{p}{n}-c=O(\sqrt{\frac{k}{n}})(=o(n^{-\frac{1}{3}}))$. If $\lambda_k>1+\sqrt{c}$, for any $1\leq i\leq k$,
\[
\frac{d_i}{\psi(\lambda_i)}\stackrel{a.s.}{\longrightarrow}1,
\]
\end{theorem}
where $\psi(\lambda_i)=\lambda_i+\frac{c\lambda_i}{\lambda_i-1}$.

To demonstrate this result, we need the following result which extends Lemma \ref{lemma:spectral}. This result can be of independent interest.
\begin{lemma}
\label{lemma:large deviation2}
For any positive semi-definite matrix $\A_n$, denoting by $l_1\geq\cdots\geq l_n$ the eigenvalues of $\A_n$, then there exist some universal constant $C>0$ and some constant $\rho$ such that the following inequality holds
\begin{eqnarray*}
&&P(||\frac{1}{n}\bSigma_{11}^{-\frac{1}{2}}\X_1^\top\A_n\X_1\bSigma_{11}^{-\frac{1}{2}}-\frac{1}{n}\tr(\A_n)\I||>t)\\
&\leq&2\exp(Ck-\frac{nt^2}{\rho\sum_{s=1}^nl_s^2/n})
\end{eqnarray*}
for all $0<t<\rho\frac{\sum_{s=1}^nl_s^2/n}{64l_1}$, where $||\cdot||$ is the spectral norm.
\end{lemma}

\begin{corollary}

For any positive semi-definite matrix $\A_n$, denoting by $l_1\geq\cdots\geq l_n$ the eigenvalues of $\A_n$, then there exist some universal constant $C>0$ and some constant $\rho$ such that the following inequality holds

\label{lemma:large deviation1}
\begin{eqnarray*}
&&P(||\frac{1}{n}\bSigma_{11}^{-\frac{1}{2}}\X_1^\top(\I+\A_n)\X_1\Sigma_{11}^{-\frac{1}{2}}-\frac{1}{n}\tr(\I+\A_n)\I||>t)\\
&\leq&2\exp(Ck-\frac{nt^2}{\rho\sum_{s=1}^n(1+l_s)^2/n}),
\end{eqnarray*}
for all $0<t<\rho\frac{\sum_{s=1}^n(1+l_s)^2/n}{64(1+l_1)}$.
\end{corollary}

Our second result shows that the spiked eigenvalues of $\S_n$, $d_1\geq\cdots\geq d_k$ are $\sqrt{n/k}$-consistent, which will be used to establish Theorem \ref{theorem:asymptotical normal}.

Compared with Theorem \ref{theorem:a.s.consistency}, we need one more condition: there exists a positive constant $C$ not depending on $n$ such that $\frac{\lambda_{i-1}}{\lambda_i}\geq C>1$, $i=1,2,\dots,k$.
This condition implies that the spiked eigenvalues are well-separated and is also used in \cite{Cai:Han:Pan:2019}.

\begin{theorem}\label{theorem:root-n/k-consistency}
Suppose that $k=o(n^{\frac{1}{4}})$, $\frac{p}{n}-c=O(\sqrt{\frac{k}{n}})(=o(n^{-\frac{3}{8}}))$, and there exists a positive constant $C$ not depending on $n$ such that $\frac{\lambda_{i-1}}{\lambda_i}\geq C>1$, $i=1,2,\dots,k$. If $\lambda_k>1+\sqrt{c}$, for any $1\leq i\leq k$,
\[
\frac{d_i-\psi(\lambda_i)}{\lambda_i}=O_p(\sqrt{\frac{k}{n}}).
\]
\end{theorem}

To demonstrate this result, we need the following Lemma \ref{lemma:diagonal}. We give some notations.

Define
\[
m_1(d)=\int\frac{x}{d-x}dF_c(x),
\]
\[
m_2(d)=\int\frac{x^2}{(d-x)^2}dF_c(x),
\]
\[
m_3(d)=\int\frac{x}{(d-x)^2}dF_c(x),
\]
\[
\H_n(d)=\frac{1}{n}\X_2(d\I -\S_{22})^{-1}\X_2^\top,
\]
\[
\bm K_n(d)=\frac{1}{n}\X_1^\top(\I+\H_n(d))\X_1,
\]
\[
\M(d)=\frac{1}{n}\bSigma_{11}^{-\frac{1}{2}}\X_1^\top(\I+\H_n(d))\X_1\bSigma_{11}^{-\frac{1}{2}}-(1+cm_1(d))\I.
\]
Let \[
\begin{array}{lll}\bSigma_{11}=\U\diag(\lambda_1,\ldots,\lambda_k)\U^\top,
\end{array}
\]
where $\U$ is an orthogonal matrix.

Obviously, \[
\bSigma_{11}^{-1}=\U\diag(\frac{1}{\lambda_1},\ldots,\frac{1}{\lambda_k})\U^\top.
\]

Also, let
\begin{eqnarray*}\A(d_i)&=&(a_{st})_{k\times k}\triangleq\U^\top\bSigma_{11}^{-\frac{1}{2}}(d_i\I-\bm K_n(d_i))\bSigma_{11}^{-\frac{1}{2}}\U\\
&=&\U^\top(\bSigma_{11}^{-1}d_i\I-\frac{1}{n}\bSigma_{11}^{-\frac{1}{2}}\X_1^\top(\I+\H_n(d_i))\X_1\bSigma_{11}^{-\frac{1}{2}})\U\\
&=&\diag(\frac{d_i-(1+cm_1(d_i))\lambda_1}{\lambda_1},\ldots,\frac{d_i-(1+cm_1(d_i))\lambda_k}{\lambda_k})-\U^\top\M(d_i)\U.
 \end{eqnarray*}

According to \cite{Bai:Yao:2008}, for $\lambda_i>1+\sqrt{c}$,  we have $m_1(\psi(\lambda_i))=\frac{1}{\lambda_i-1}$.
Thus, for the $j$-th diagonal element  with index $j\neq i$,
\begin{eqnarray*}e_j\triangleq\frac{d_i-(1+cm_1(d_i))\lambda_j}{\lambda_j}&=&\frac{d_i}{\lambda_j}-(1+cm_1(d_i))\\
&\rightarrow&\frac{\psi(\lambda_i)}{\lambda_j}-(1+cm_1(\psi(\lambda_i)))\\
&=&(\frac{\lambda_i}{\lambda_j}-1)(1+\frac{c}{\lambda_i-1})
\end{eqnarray*}
is uniformly bounded away from $0$.

Let
\begin{equation}
\label{eq:determinant}
\det(\A(d_i))=\prod_{j=1}^k\hat{e}_j\triangleq\prod_{j=1}^k(e_j+\epsilon_j),
\end{equation}
where $\hat{e}_j$'s are eigenvalues of matrix $\A(d_i)$.

Similar to the proof of \eqref{thm1:eq9} in Theorem \ref{theorem:a.s.consistency}, we have
\[
\sum_{i=1}^k|\epsilon_j|=O_p(\sqrt{\frac{k^3}{n}}).
\]

For $j\neq i$, under the condition that $k=o(n^{\frac{1}{3}})$, $\hat{e}_j$ is uniformly bounded away from $0$ since the contributions from $\U^\top\M(d_i)\U$ tend to $O_p(\sqrt{\frac{k^3}{n}})=o_p(1)$.

By \eqref{eq:determinant}, for sufficiently large $n$, $\det(\A(d_i))=0$ if and only if $\hat{e}_i=0$.

However, $\hat{e}_i=0$ does not imply that $a_{ii}=0$.

For any $1\leq i\leq k$, the following lemma gives the order of the $i$-th diagonal element of matrix $\A(d_i)$,  which is the key to establish Theorems \ref{theorem:root-n/k-consistency} and \ref{theorem:asymptotical normal}.

\begin{lemma}
\label{lemma:diagonal}
Suppose that $k=o(n^{\frac{1}{4}})$, $\frac{p}{n}-c=O(\sqrt{\frac{k}{n}})(=o(n^{-\frac{3}{8}}))$, and there exists a positive constant $C$ not depending on $n$ such that $\frac{\lambda_{i-1}}{\lambda_i}\geq C>1$, $i=1,2,\dots,k$. If $\lambda_k>1+\sqrt{c}$, for any $1\leq i\leq k$,
\[
a_{ii}=\frac{d_i-(1+cm_1(d_i))\lambda_i}{\lambda_i}-[\U^\top\M(d_i)\U]_{ii}=O_p(\frac{k^2}{n}).
\]
\end{lemma}

Our third result demonstrates that the spiked eigenvalues of $\S_n$, $d_1\geq\cdots\geq d_k$ are asymptotical normal.
\begin{theorem}\label{theorem:asymptotical normal}
Suppose that $k=o(n^{\frac{1}{4}})$, $\frac{p}{n}-c=o(n^{-\frac{1}{2}})$, and there exists a positive constant $C$ not depending on $n$ such that $\frac{\lambda_{i-1}}{\lambda_i}\geq C>1$, $i=1,2,\dots,k$. If $\lambda_k>1+\sqrt{c}$, for any $1\leq i\leq k$,
\[
\frac{\sqrt{n}(d_i-\psi(\lambda_i))}{\lambda_i}\stackrel{a.s.}{\longrightarrow}T(\psi(\lambda_i)),
\]
where
\[
T(\psi(\lambda_i))\triangleq(1+cm_3(\psi(\lambda_i))\lambda_i)^{-1}[\U^\top\bSigma_{11}^{-\frac{1}{2}}\bm R_n(\psi(\lambda_i))\bSigma_{11}^{-\frac{1}{2}}\U]_{ii},
\]
and
\[
\bm R_n(\psi(\lambda_i))=\frac{1}{\sqrt{n}}[\X_1^\top(\I+\H_n(\psi(\lambda_i)))\X_1-\tr(\I+\H_n(\psi(\lambda_i)))\bSigma_{11}].
\]
\end{theorem}
For any $1\leq i\leq k$, the limiting distribution of $T(\psi(\lambda_i))$ is normal. For more details, we refer to \cite{Bai:Yao:2008} and \cite{Li:Han:Yao:2019}.

Especially, we consider the case $\bSigma_{11}$ is diagonal as discussed in \cite{Paul:2007}. Similar to the discussion in \cite{Bai:Yao:2008}, for any $1\leq i\leq k$, we have
\[
\frac{\sqrt{n}(d_i-\psi(\lambda_i))}{\lambda_i}\stackrel{d}{\longrightarrow}N(0,\sigma_{\lambda_i}^2),
\]
where $\sigma_{\lambda_i}^2=2-\frac{2c}{(\lambda_i-1)^2}$.

Note that throughout this paper we assume that $\x(t)$ is normal. Under the condition that for any $1\leq i\leq n$, $1\leq j\leq k$, $E((\bSigma_{11}^{-\frac{1}{2}}\x_1(t_i))_j^4)\leq C$, we have the following results.

\begin{lemma}
\label{corollary:large deviation2}
For any positive semi-definite matrix $\A_n$, denoting by $l_1\geq\cdots\geq l_n$ the eigenvalues of $\A_n$, suppose that for any $1\leq i\leq n$, $1\leq j\leq k$, $E((\bSigma_{11}^{-\frac{1}{2}}\x_1(t_i))_j^4)\leq C$ hold,  we have
\[
||\frac{1}{n}\bSigma_{11}^{-\frac{1}{2}}\X_1^\top\A_n\X_1\bSigma_{11}^{-\frac{1}{2}}-\frac{1}{n}\tr(\A_n)I||=\sqrt{\frac{1}{n}\sum_{s=1}^nl_s^2}O_p(\frac{k}{\sqrt{n}}).
\]
\end{lemma}

\begin{corollary}
\label{corollary:large deviation1}
For any positive semi-definite matrix $\A_n$, denoting by $l_1\geq\cdots\geq l_n$ the eigenvalues of $\A_n$, suppose that for any $1\leq i\leq n$, $1\leq j\leq k$, $E((\bSigma_{11}^{-\frac{1}{2}}\x_1(t_i))_j^4)\leq C$ hold,  we have
\[
||\frac{1}{n}\bSigma_{11}^{-\frac{1}{2}}\X_1^\top(\I+\A_n)\X_1\bSigma_{11}^{-\frac{1}{2}}-\frac{1}{n}\tr(\I+\A_n)I||=\sqrt{\frac{1}{n}\sum_{s=1}^n(1+l_s)^2}O_p(\frac{k}{\sqrt{n}}).
\]

\end{corollary}

\begin{corollary}\label{corollary:a.s.consistency}
Suppose that $k=o(n^{\frac{1}{4}})$, $\frac{p}{n}-c=O(\frac{k}{\sqrt{n}})(=o(n^{-\frac{1}{4}}))$, and for any $1\leq i\leq n$, $1\leq j\leq k$, $E((\bSigma_{11}^{-\frac{1}{2}}\x_1(t_i))_j^4)\leq C$. If $\lambda_k>1+\sqrt{c}$, for any $1\leq i\leq k$,
\[
\frac{d_i}{\psi(\lambda_i)}\stackrel{a.s.}{\longrightarrow}1,
\]
\end{corollary}
where $\psi(\lambda_i)=\lambda_i+\frac{c\lambda_i}{\lambda_i-1}$.

\begin{corollary}\label{corollary:root-n/k-consistency}
Suppose that $k=o(n^{\frac{1}{5}})$, $\frac{p}{n}-c=O(\frac{k}{\sqrt{n}})(=o(n^{-\frac{3}{10}}))$, for any $1\leq i\leq n$, $1\leq j\leq k$, $E((\bSigma_{11}^{-\frac{1}{2}}\x_1(t_i))_j^4)\leq C$, and there exists a positive constant $C$ not depending on $n$ such that $\frac{\lambda_{i-1}}{\lambda_i}\geq C>1$, $i=1,2,\dots,k$. If $\lambda_k>1+\sqrt{c}$, for any $1\leq i\leq k$,
\[
\frac{d_i-\psi(\lambda_i)}{\lambda_i}=O_p(\frac{k}{\sqrt{n}}).
\]
\end{corollary}

\begin{lemma}
\label{corollary:diagonal}
Suppose that $k=o(n^{\frac{1}{5}})$, $\frac{p}{n}-c=O(\frac{k}{\sqrt{n}})(=o(n^{-\frac{3}{10}}))$, for any $1\leq i\leq n$, $1\leq j\leq k$, $E((\bSigma_{11}^{-\frac{1}{2}}\x_1(t_i))_j^4)\leq C$, and there exists a positive constant $C$ not depending on $n$ such that $\frac{\lambda_{i-1}}{\lambda_i}\geq C>1$, $i=1,2,\dots,k$. If $\lambda_k>1+\sqrt{c}$, for any $1\leq i\leq k$,
\[
a_{ii}=\frac{d_i-(1+cm_1(d_i))\lambda_i}{\lambda_i}-[\U^\top\M(d_i)\U]_{ii}=O_p(\frac{k^3}{n}).
\]
\end{lemma}

\begin{corollary}\label{corollary:asymptotical normal}
Suppose that $k=o(n^{\frac{1}{6}})$, $\frac{p}{n}-c=o(n^{-\frac{1}{2}})$, for any $1\leq i\leq n$, $1\leq j\leq k$, $E((\bSigma_{11}^{-\frac{1}{2}}\x_1(t_i))_j^4)\leq C$, and there exists a positive constant $C$ not depending on $n$ such that $\frac{\lambda_{i-1}}{\lambda_i}\geq C>1$, $i=1,2,\dots,k$. If $\lambda_k>1+\sqrt{c}$, for any $1\leq i\leq k$,
\[
\frac{\sqrt{n}(d_i-\psi(\lambda_i))}{\lambda_i}\stackrel{a.s.}{\longrightarrow}(1+cm_3(\psi(\lambda_i))\lambda_i)^{-1}[\U^\top\bSigma_{11}^{-\frac{1}{2}}\bm R_n(\psi(\lambda_i))\Sigma_{11}^{-\frac{1}{2}}\U]_{ii},
\]
where
\[
\bm R_n(\psi(\lambda_i))=\frac{1}{\sqrt{n}}[\X_1^\top(\I+\H_n(\psi(\lambda_i))X_1-\tr(\I+\H_n(\psi(\lambda_i))\bSigma_{11}].
\]
\end{corollary}

\subsection{Extended AIC}
In this subsection, we extend our results to the case $n,p\to\infty$, with $p/n\to c>0$. In this case, the AIC tends to underestimate the number of signals $k$ (see Section \ref{section:experiments}). It is well-known that, for fixed $p$, the AIC tends to overestimate the number of signals $k$.  This implies that, in the case $n,p\to\infty$, with $p/n\to c>0$, the AIC does not work and should be re-defined. In view of this, we propose an extended AIC (EAIC).

The EAIC estimates $k$ by maximizing $\ell(k')$:
\begin{equation}
\label{eq:aic5}
\hat{k}=\arg\max_{k'}\ell(k'),
\end{equation}
where
\begin{equation}
\label{eq:aic6}\ell(k')=-\frac{1}{2}n(\sum_{i=1}^{k'}\log d_i+(p-k')\log\hat{\lambda}_{k'})-\gamma k'(p-k'/2+1/2),
\end{equation}
where $\gamma=\varphi(c)=1/2+\sqrt{1/c}-\log(1+\sqrt{c})/c$.

 Note that when $\gamma$ is equal to $1$, (\ref{eq:aic5}) is the AIC. In the following, we first show that the EAIC is a natural extension of the AIC for fixed $p$. In the next subsection, we further explain that, for the EAIC, why $\gamma$ should be defined as $\varphi(c)=1/2+\sqrt{1/c}-\log(1+\sqrt{c})/c$.

In the following two cases,
\begin{enumerate}
\item[(1)]$p$ fixed, $n\to\infty$,
\item[(2)] $n,p\to\infty$ with $p/n\to 0$,
\end{enumerate} $c$ is defined as $c=p/n$. Noting that $c\to 0+$, by Taylor's expansion, we get
\[
\begin{array}{lll}\varphi(c)&=&1/2+\sqrt{1/c}-\log(1+\sqrt{c})/c\\
&=&1/2+\sqrt{1/c}-(\sqrt{c}-\frac{c}{2}(1+o(1)))/c\\
&=&1+o(1)\\
&\rightarrow&1.
\end{array}
\]

As a result, in the following three cases,
\begin{enumerate}
\item[(1)] $p$ fixed, $n\to\infty$,
\item[(2)] $n,p\to\infty$ with $p/n\to 0$,
\item[(3)] $n,p\to\infty$ with $p/n\to c$, where $0<c<\infty$,
\end{enumerate}
approximately, the AIC may be uniformly written as (\ref{eq:aic5}).

Note that the EAIC is different from that of \cite{Bai:Choi:Fujikoshi:2018} which proposes the following BCF for estimating the number of principal components $k$.

In the case $0<c<1$, \citeauthor{Bai:Choi:Fujikoshi:2018} \cite{Bai:Choi:Fujikoshi:2018} estimates $k$ by minimizing $\ell^1(k')$:
\begin{equation}
\label{eq:aic1}
\hat{k}=\arg\min_{k'}\ell^1(k'),
\end{equation}
where $\ell^1(k')=(p-k')\log \bar{d}_{k'}-\sum_{i=k'+1}^p\log d_i-(p-k'-1)(p-k'+2)/n$, $\bar{d}_{k'}=\sum_{i=k'+1}^pd_i/(p-k')$.

Note that,  the BCF defined in (\ref{eq:aic1}) is essentially equivalent to the AIC and is referred as the AIC in \cite{Bai:Choi:Fujikoshi:2018}. In the next subsection, we will show that,  at low SNR, the BCF defined in (\ref{eq:aic1}) tends to underestimate the number of signals $k$, especially in the case $c\rightarrow1-$  (see also Section \ref{section:experiments}).

In the case $c>1$, \citeauthor{Bai:Choi:Fujikoshi:2018} \cite{Bai:Choi:Fujikoshi:2018} replaces the $p$ in (\ref{eq:aic1}) by $n-1$
 and estimates $k$ by minimizing $\ell^2(k')$:
\begin{equation}
\label{eq:aic2}
\hat{k}=\arg\min_{k'}\ell^2(k'),
\end{equation}
where $\ell^2(k')=(n-1-k')\log \bar{d}_{k'}-\sum_{i=k'+1}^{n-1}\log d_i-(n-k'-2)(n-k'+1)/p$, $\bar{d}_{k'}=\sum_{i=k'+1}^{n-1}d_i/(n-1-k')$.

Simulations show that, the BCF defined in (\ref{eq:aic2}) is better than the AIC and is referred as the quasi-AIC in \cite{Bai:Choi:Fujikoshi:2018}. However, at low SNR, the BCF defined in (\ref{eq:aic2}) still tends to underestimate the number of signals $k$, especially in the case $c\rightarrow1+$  (see also Section \ref{section:experiments}).

\subsection{Consistency of the EAIC-type method}

Noting that the EAIC tends to overestimate the number of signals $k$, we propose the EAIC-type method which chooses the correct $k$ with probability
tending to one when $n$ goes to infinity.

The EAIC-type method estimates $k$ by maximizing $\ell(k')$:
\begin{equation}
\label{eq:aic}
\hat{k}=\arg\max_{k'}\ell(k'),
\end{equation}
where
\begin{equation}
\label{eq:aic3}\ell(k')=-\frac{1}{2}n(\sum_{i=1}^{k'}\log d_i+(p-k')\log\hat{\lambda}_{k'})-\gamma k'(p-k'/2+1/2),
\end{equation}
where $\gamma>\varphi(c)=1/2+\sqrt{1/c}-\log(1+\sqrt{c})/c$.

To achieve the consistency of \eqref{eq:aic}, we  need two additional conditions,
\begin{equation}
\label{eq:consistency1}
\psi(\lambda_k)-1-\log \psi(\lambda_k)>2\gamma c,
\end{equation}
\begin{equation}
\label{eq:consistency2}
\lambda_k>1+\sqrt{c},
\end{equation}
where $\psi(\lambda_k)=\lambda_k+\frac{c\lambda_k}{\lambda_k-1}$.

Since there is a tuning parameter $\gamma$ in (\ref{eq:consistency1}), $\lambda_k>1+\sqrt{c}$ does not implies $\psi(\lambda_k)-1-\log \psi(\lambda_k)>2\gamma c$. However, by $\gamma>\varphi(c)$ and $\psi(\lambda_k)-1-\log \psi(\lambda_k)>2\gamma c$, we have $\lambda_k>1+\sqrt{c}$.

Next, we establish the consistency of \eqref{eq:aic}.

\begin{theorem}
\label{theorem:4}  In the case $n,p\to\infty$ with $p/n\to c>0$, let $\ell(k')$ be the penalized likelihood function defined in (\ref{eq:aic3}). Suppose that $k=o(n^{\frac{1}{3}})$ and $\frac{p}{n}-c=O(\sqrt{\frac{k}{n}})(=o(n^{-\frac{1}{3}}))$, and the number of candidate models, $q$, satisfies $q = o(p)$.\\
For $k'<k$, if (\ref{eq:consistency1}) and (\ref{eq:consistency2}) hold (or if $\gamma>\varphi(c)$ and (\ref{eq:consistency1}) hold),
\[
P(\ell(k)>\ell(k'))\rightarrow1.
\]
For $k'>k$, if $\gamma>\varphi(c)$,
\[
P(\ell(k)>\ell(k'))\rightarrow1.
\]
where $\varphi(c)=1/2+\sqrt{1/c}-\log(1+\sqrt{c})/c$.
\end{theorem}

Note that throughout this paper, we assume that $\x(t)$ is normal. Under the more general condition that for any $1\leq i\leq n$, $1\leq j\leq k$, $E((\bSigma_{11}^{-\frac{1}{2}}\x_1(t_i))_j^4)\leq C$, we have the following result.
\begin{corollary}
\label{corollary:consistency}  In the case $n,p\to\infty$ with $p/n\to c>0$, let $\ell(k')$ be the penalized likelihood function defined in (\ref{eq:aic3}). Suppose that $k=o(n^{\frac{1}{4}})$, $\frac{p}{n}-c=O(\frac{k}{\sqrt{n}})(=o(n^{-\frac{1}{4}}))$, for any $1\leq i\leq n$, $1\leq j\leq k$, $E((\bSigma_{11}^{-\frac{1}{2}}\x_1(t_i))_j^4)\leq C$, and the number of candidate models, $q$, satisfies $q = o(p)$.\\
For $k'<k$, if (\ref{eq:consistency1}) and (\ref{eq:consistency2}) hold (or if $\gamma>\varphi(c)$ and (\ref{eq:consistency1}) hold),
\[
P(\ell(k)>\ell(k'))\rightarrow1.
\]
For $k'>k$, if $\gamma>\varphi(c)$,
\[
P(\ell(k)>\ell(k'))\rightarrow1.
\]
where $\varphi(c)=1/2+\sqrt{1/c}-\log(1+\sqrt{c})/c$.
\end{corollary}

Note that for fixed $p$ and $n\rightarrow\infty$, $\gamma>\varphi(c)$ becomes $\gamma>1$. It is well-known that, for fixed $p$, the AIC tends to overestimate the number of signals $k$. To achieve the consistency of the AIC-type method, the tuning parameter $\gamma$ should be larger than one. That is, in the case $n,p\to\infty$ with $p/n\to c>0$, the EAIC can be seen as a natural extension of the AIC for fixed $p$. There is a similar discussion in the case $n,p\to\infty$ with $p/n\to 0$.

\begin{remark}
\label{eq:remark2} In the following two cases,
\begin{enumerate}
\item[(1)] $p$ fixed, $n\to\infty$,
\item[(2)] $n,p\to\infty$ with $p/n\to 0$,
\end{enumerate} if the AIC is defined as the degeneration of the EAIC in the case $n,p\to\infty$ with $p/n\to c>0$, i.e., $\gamma=\lim_{c\rightarrow 0+}\varphi(c)=1$, then we have essentially demonstrated that, the AIC tends to overestimate $k$. To achieve the consistency of the AIC-type method in the above two cases, $\gamma>1$ is required.
\end{remark}

\begin{remark}
\label{eq:remark3}
By Theorem \ref{theorem:4}, in the case $n,p\to\infty$ with $p/n\to c>0$, we have actually explained why the AIC tends to underestimate the number of signals $k$.
We note that $\gamma>\varphi(c)$ is required in our theoretic analysis, to avoid overestimating the number of signals $k$. On the other hand, (\ref{eq:consistency1}) requires that $\gamma$ cannot be too large, to avoid underestimating the number of signals $k$. Noting that, in the case $n,p\to\infty$ with $p/n\to c>0$, we always have $\varphi(c)<1$. Thus, the penalty of the AIC may be too large. As a result, it is not surprising that the AIC may tend to underestimate the number of signals $k$.
\end{remark}

Since $\gamma>\varphi(c)$ is required in our theoretic analysis, we set $\gamma=1.1\varphi(c)$ for simulation studies which gives good performance. In this sense, the EAIC-type method is essentially tuning-free.

For fixed $k$, to prove the consistency of the BCF, \citeauthor{Bai:Choi:Fujikoshi:2018} \cite{Bai:Choi:Fujikoshi:2018} gives the following two conditions,
\begin{enumerate}
\item[(1)] in the case $0<c<1$,
\begin{equation}
\label{eq:consistency3}
\psi(\lambda_k)-1-\log \psi(\lambda_k)>2c \,\,\,\textrm{and}
\end{equation}
\item[(2)] in the case $c>1$,
\begin{equation}
\label{eq:consistency4}
\psi(\lambda_k)/c-1-\log(\psi(\lambda_k)/c)>2/c.
\end{equation}
\end{enumerate}

Compared with (\ref{eq:consistency3}) and (\ref{eq:consistency4}), (\ref{eq:consistency1}) contains a tuning parameter $\gamma$. By simulations, we found that when $c\geq0.12$, we have $\gamma=1.1\varphi(c)<1$. That is, when $0.12\leq c\leq1$, (\ref{eq:consistency1}) is weaker than (\ref{eq:consistency3}). This implies that, when $0.12\leq c\leq1$, in the case where (\ref{eq:consistency1}) holds while (\ref{eq:consistency3}) does not hold, the EAIC-type method is better than the BCF. When $ c>1$, although it is difficult to compare (\ref{eq:consistency1}) with (\ref{eq:consistency4}), by simulations, we found that there do exist cases where (\ref{eq:consistency1}) holds while (\ref{eq:consistency4}) does not hold (see Section \ref{section:experiments}). This implies that, in the case  where (\ref{eq:consistency1}) holds while (\ref{eq:consistency4}) does not hold, the EAIC-type method is still better than the BCF. We also note that in the case $c>1$, if $c$ is close to one, (\ref{eq:consistency4}) and (\ref{eq:consistency3}) are almost the same. Thus, in the case where $c$ is close to one, the EAIC-type method outperforms the BCF. The reason lies in that, when $c=p/n$ is close to one, (\ref{eq:consistency1}) is weaker than (\ref{eq:consistency3}) and (\ref{eq:consistency4}). As a result, the EAIC-type method outperforms the BCF at least in the following two cases,
\begin{enumerate}
\item[(1)] in the case $0.12\leq c\leq1$,
\item[(2)] in the case $c>1$, and $c$ is close to one.
\end{enumerate}
Moreover, note that (\ref{eq:aic1}) is essentially equivalent to the AIC. Since when $c\geq0.12$, we have $\gamma=1.1\varphi(c)<1$. That is, when (1) $0.12\leq c\leq 1$; (2) $c$ is larger than one and is close to one, the BCF tends to underestimate the number of signals $k$.

On the other hand, in the case where $c$ is larger than one, simulations show that the EAIC-type method is still comparable to the BCF (see Section \ref{section:experiments}).

\begin{remark}
\label{eq:remark4}
For fixed k, \citeauthor{Bai:Choi:Fujikoshi:2018} \cite{Bai:Choi:Fujikoshi:2018}  established the consistency of the BCF. Along the way of the proposed method in this paper, it is possible to extend their results to increasing $k$.
\end{remark}

\begin{remark}
\label{eq:remark5}
Although the BCF  defined in (\ref{eq:aic1}) and (\ref{eq:aic2}) is tuning-free and simulation results are encouraging, \citeauthor{Bai:Choi:Fujikoshi:2018} \cite{Bai:Choi:Fujikoshi:2018} essentially defines two different criteria in the case $0<c<1$ and $c>1$, respectively. As a result, to achieve the consistency of the BCF, two different consistency conditions are required (i.e., (\ref{eq:consistency3}) and (\ref{eq:consistency4})). Compared with (\ref{eq:aic1}) and (\ref{eq:aic2}), (\ref{eq:aic}) with $\gamma>\varphi(c)$ is a natural extension of the AIC for fixed $p$. We also note that, both the formula (\ref{eq:aic}) and the consistency condition (\ref{eq:consistency1}) are more simple.
\end{remark}

\begin{table}[h]
\caption{performance of ILP, BIC, AIC, Modified AIC, KN: $p=12$, $k=3$, $\lambda=1$, $SNR=\delta\sqrt{4(p-k/2+1/2)\log\log n/n}$}
\label{tab:1a}
\centering
\scalebox{1}{
\begin{tabular}{ccc|cc|cc|cc|ccccccccc}
\\
\hline
&\multicolumn{ 2}{c|}{$\delta=1.5$}&\multicolumn{ 2}{c|}{$\delta=1.75$}&\multicolumn{ 2}{c|}{$\delta=2$}& \multicolumn{ 2}{c|}{$\delta=2.25$}&\multicolumn{ 2}{c}{$\delta=2.5$}\\
\hline
&                           Prob & Mean &      Prob & Mean & Prob & Mean  &    Prob & Mean &   Prob & Mean &\\
\hline
$n=100$&                            &  &       &  &  &   &     &  &    &  &\\
\hline
$ILP^{\frac{1}{2}}$($\gamma=1.00$) &   0.87  & 2.98&     0.92& 3.03&     0.94  & 3.05&0.95  & 3.06 &     0.95& 3.06\\
ILP ($\gamma=1.00$)      &   0.79  & 2.82&     0.93& 2.95&     0.96  & 2.98&0.97  & 3.00 &     0.98& 3.01\\
BIC                     &   0.43  & 2.42&     0.61& 2.60&     0.77  & 2.77&0.87  & 2.87 &     0.92& 2.92\\
AIC                    &   0.84  & 3.13&     0.86& 3.12&     0.87  & 3.15&0.86  & 3.16 &     0.86& 3.16\\
Modified AIC           &   0.56  & 2.56&     0.67& 2.67&     0.84  & 2.84&0.93  & 2.93 &     0.96& 2.97\\
KN                   &   0.41  & 2.41&     0.60& 2.59&     0.73  & 2.73&0.86  & 2.89 &     0.92& 2.92\\
\hline
&                            &  &       &  &  &   &     &  &    &  &\\
$n=200$&                            &  &       &  &  &   &     &  &    &  &\\
\hline
$ILP^{\frac{1}{2}}$($\gamma=1.00$)                    &   0.96  & 3.01&     0.98& 3.02&     0.98  & 3.02&0.98  & 3.02 &     0.98& 3.02\\
ILP ($\gamma=1.00$)                    &   0.97  & 2.97&     1.00& 3.00&     1.00  & 3.00&1.00  & 3.00 &     1.00& 3.00\\
BIC                    &   0.47  & 2.47&     0.76& 2.76&     0.90  & 2.90&0.97  & 2.97 &     1.00& 3.00\\
AIC                    &   0.91  & 3.10&     0.90& 3.12&     0.91 & 3.10&0.86  & 3.16 &     0.86& 3.16\\
Modified AIC           &   0.80  & 2.80&     0.94& 2.94&     1.00  & 3.00&1.00  & 3.00 &     1.00& 3.00\\
KN                   &   0.59  & 2.59&     0.84& 2.84&     0.96  & 2.96&0.99  & 2.99 &     1.00& 3.00\\
\hline
&                            &  &       &  &  &   &     &  &    &  &\\
$n=500$&                            &  &       &  &  &   &     &  &    &  &\\
\hline
$ILP^{\frac{1}{2}}$($\gamma=1.00$) &   0.97  & 3.02&     0.98& 3.02&     0.98  & 3.02&0.98  & 3.02 &     0.98& 3.02\\
ILP ($\gamma=1.00$)       &   0.96  & 2.96&     1.00& 3.00&     1.00  & 3.00&1.00  & 3.00 &     1.00& 3.00\\
BIC                    &   0.48  & 2.48&     0.80& 2.80&     0.94  & 2.94&1.00  & 3.00 &     1.00& 3.00\\
AIC                    &   0.86  & 3.15&     0.86& 3.16&     0.85  & 3.16&0.84  & 3.19 &     0.84& 3.19\\
Modified AIC          &   0.92  & 2.92&     0.98& 2.98&     1.00  & 3.00&1.00  & 3.00 &     1.00& 3.00\\
KN                   &   0.81  & 2.81&     0.95& 2.95&     0.99  & 2.99&0.99  & 2.99 &     1.00& 3.00\\
\hline
&                            &  &       &  &  &   &     &  &    &  &\\
$n=800$&                            &  &       &  &  &   &     &  &    &  &\\
\hline
$ILP^{\frac{1}{2}}$($\gamma=1.00$) &   0.98  & 3.03&     0.99& 3.01&     0.99  & 3.01&0.99  & 3.01 &     0.99& 3.01\\
ILP ($\gamma=1.00$)      &   0.99  & 2.99&     1.00& 3.00&     1.00  & 3.00&1.00  & 3.00 &     1.00& 3.00\\
BIC                    &   0.59  & 2.59&     0.90& 2.90&     0.95  & 2.95&1.00  & 3.00 &     1.00& 3.00\\
AIC                    &   0.84  & 3.18&     0.84& 3.18&     0.84  & 3.18&0.84  & 3.17 &     0.84& 3.17\\
Modified AIC           &   0.95  & 2.95&     1.00& 3.00&     1.00  & 3.00&1.00  & 3.00 &     1.00& 3.00\\
KN                   &   0.84  & 2.84&     1.00& 3.00&     1.00  & 3.00&1.00  & 1.00 &     1.00& 3.00\\
\hline

\end{tabular}
}
\end{table}

\begin{table}[h]
\caption{performance of ILP, AIC, Modified AIC, EAIC-type, BCF, KN: $n=500$, $p=200$, $k=10$, $\lambda=1$, $SNR=\delta$}
\label{tab:5a}
\centering
\scalebox{1}{
\begin{tabular}{ccc|cc|cc|cc|ccccccccc}
\\
\hline
& \multicolumn{ 2}{c|}{$\delta=0.5$}&\multicolumn{ 2}{c|}{$\delta=1$}&\multicolumn{ 2}{c|}{$\delta=1.5$}&\multicolumn{ 2}{c|}{$\delta=2$}&\multicolumn{ 2}{c}{$\delta=2.5$}\\
\hline
&                           Prob & Mean &      Prob & Mean & Prob & Mean  &    Prob & Mean &   Prob & Mean &\\
\hline
ILP ($\gamma=1.00$)       &0.00  & 0.00 &     0.00& 1.51&   0.00  & 8.39&     0.10& 9.10&     0.85  & 9.85\\
AIC                       &0.00  & 3.07 &     0.24& 9.23&   1.00  & 10.0&     1.00& 10.0&     1.00  & 10.0\\
Modified AIC              &0.00  & 0.00 &     0.00& 5.26&   0.00  & 7.44&     0.02& 9.02&     0.67  & 9.67&\\
EAIC-type                 &0.00  & 4.40 &     0.52& 9.52&   1.00  & 10.0&     1.00& 10.0&     1.00  & 10.0&\\
BCF                       &0.00  & 3.07 &     0.24& 9.23&   1.00  & 10.0&     1.00& 10.0&     1.00  & 10.0&\\
KN                        &0.00  & 1.81 &     0.02& 8.94&   0.89  & 9.89&     1.00& 10.0&     1.00  & 10.0&\\

\hline

\end{tabular}
}
\end{table}

\begin{table}[h]
\caption{performance of ILP, AIC, Modified AIC, EAIC-type, BCF, KN: $n=200$, $p=500$, $k=10$, $\lambda=1$, $SNR=\delta$}
\label{tab:5b}
\centering
\scalebox{1}{
\begin{tabular}{ccc|cc|cc|cc|ccccccccc}
\\
\hline
& \multicolumn{ 2}{c|}{$\delta=1.5$}&\multicolumn{ 2}{c|}{$\delta=2.5$}&\multicolumn{ 2}{c|}{$\delta=2.68$}&\multicolumn{ 2}{c|}{$\delta=3.5$}&\multicolumn{ 2}{c}{$\delta=4.5$}\\
\hline
&                           Prob & Mean &      Prob & Mean &Prob & Mean&    Prob & Mean  &    Prob & Mean &   \\
\hline
ILP ($\gamma=1.00$)       &0.00  & 0.00 &     0.00& 0.00&   0.00  & 1.00&   0.00  & 1.34&     0.00& 5.34&     \\
AIC                       &0.00  & 1.16 &     0.00& 7.05&   0.00  & 8.00&   0.12  & 9.10&     0.70& 9.70&     \\
Modified AIC              &0.00  & 0.00 &     0.00& 0.00&   0.00  & 0.00&   0.00  & 0.00&     0.00& 1.42&     \\
EAIC-type                     &0.00  & 5.60 &     0.31& 9.31&   0.43  & 9.43&   0.92  & 9.92&     1.00& 10.0&     \\
BCF                       &0.00  & 3.75 &     0.11& 9.04&   0.25  & 9.22&   0.84  & 9.84&     1.00& 10.0&     \\
KN                        &0.00  & 3.62 &     0.01& 8.43&   0.07  & 9.06&   0.45  & 9.45&     0.94& 9.94&     \\

\hline

\end{tabular}
}
\end{table}

\begin{table}[h]
\caption{performance of ILP, AIC, Modified AIC, EAIC-type, BCF, KN: $n=200$, $p=200$, $k=10$, $\lambda=1$, $SNR=\delta$}
\label{tab:5c}
\centering
\scalebox{1}{
\begin{tabular}{ccc|cc|cc|cc|ccccccccc}
\\
\hline
& \multicolumn{ 2}{c|}{$\delta=1$}&\multicolumn{ 2}{c|}{$\delta=1.5$}&\multicolumn{ 2}{c|}{$\delta=2$}&\multicolumn{ 2}{c|}{$\delta=2.5$}&\multicolumn{ 2}{c}{$\delta=3$}\\
\hline
&                           Prob & Mean &      Prob & Mean & Prob & Mean  &    Prob & Mean &   Prob & Mean &\\
\hline
ILP ($\gamma=1.00$)       &0.00  & 0.00 &     0.00& 0.00&   0.00  & 3.22&     0.00& 6.49&     0.00  & 8.59&\\
AIC                       &0.00  & 3.73 &     0.00& 8.10&   0.32  & 9.32&     0.88& 9.88&     0.98  & 9.98\\
Modified AIC              &0.00  & 3.73 &     0.00& 1.00&   0.00  & 1.08&     0.00& 3.23&     0.00  & 6.63&\\
EAIC-type                 &0.00   & 6.37 &     0.25& 9.17&   0.85  & 9.85&     0.94& 10.0&     0.97  & 10.0&\\
BCF                      &0.00   & 3.73 &     0.00& 8.10&   0.32  & 9.32&     0.88& 9.88&     0.98  & 9.98&\\
KN                       &0.00  & 3.04 &      0.00& 7.13&   0.09  & 8.97&     0.58& 9.58&     0.92  & 9.92&\\

\hline

\end{tabular}
}
\end{table}

\begin{table}[h]
\caption{performance of EAIC-type: $\gamma=1.1\varphi(c)$, $k=10$, $\lambda=1$, $SNR=2\sqrt{p/n}$}
\label{tab:5d}
\centering
\scalebox{1}{
\begin{tabular}{ccc|cc|cc|cc|ccccccccc}
\\
\hline
& \multicolumn{ 2}{c|}{$n=100$}&\multicolumn{ 2}{c|}{$n=200$}&\multicolumn{ 2}{c|}{$n=300$}&\multicolumn{ 2}{c|}{$n=400$}&\multicolumn{ 2}{c}{$n=500$}\\
\hline
&                              Prob & Mean &      Prob & Mean & Prob & Mean  &    Prob & Mean &   Prob & Mean &\\
\hline
$p=100$                       &\textbf{0.55}  & \textbf{9.76} &     0.82& 9.90&   0.85  & 9.89&     0.93& 9.97&     0.90  & 9.94\\
$p=200$                       &0.65  & 9.89 &     \textbf{0.85}& \textbf{9.85}&   0.89  & 9.91&     0.93& 9.98&     0.95  & 10.0\\
$p=300$                       &0.66  & 9.96 &     0.88& 9.89&   \textbf{0.91}  & \textbf{9.91}&     0.96& 9.96&     0.93  & 9.93&\\
$p=400$                       &0.63   & 10.0 &     0.78& 9.78&   0.92  & 9.92&     \textbf{0.97}& \textbf{9.97}&     0.94  & 9.94&\\
$p=500$                       &0.63   & 10.1 &     0.84& 9.84&   0.89  & 9.89&     0.94& 9.94&     \textbf{0.97}  & \textbf{9.97}&\\

\hline

\end{tabular}
}
\end{table}

\begin{table}[h]
\caption{performance of BCF : $k=10$, $\lambda=1$, $SNR=2\sqrt{p/n}$}
\label{tab:5e}
\centering
\scalebox{1}{
\begin{tabular}{ccc|cc|cc|cc|ccccccccc}
\\
\hline
& \multicolumn{ 2}{c|}{$n=100$}&\multicolumn{ 2}{c|}{$n=200$}&\multicolumn{ 2}{c|}{$n=300$}&\multicolumn{ 2}{c|}{$n=400$}&\multicolumn{ 2}{c}{$n=500$}\\
\hline
&                           Prob & Mean &      Prob & Mean & Prob & Mean  &    Prob & Mean &   Prob & Mean &\\
\hline
$p=100$                       &\textbf{0.30}  & \textbf{9.13} &     0.63& 9.63&   0.74  & 9.74&     0.91& 9.91&     0.89  & 9.93\\
$p=200$                       &0.46  & 9.39 &     \textbf{0.32}& \textbf{9.32}&   0.54  & 9.54&     0.85& 9.85&     0.84  & 9.84\\
$p=300$                       &0.46  & 9.46 &     0.31& 9.31&   \textbf{0.34}  & \textbf{9.34}&     0.63& 9.63&     0.75  & 9.75&\\
$p=400$                       &0.47   & 9.47 &     0.53& 9.53&   0.51  & 9.51&     \textbf{0.45}& \textbf{9.45}&     0.74  & 9.74&\\
$p=500$                       &0.53   & 9.53 &     0.52& 9.52&   0.64  & 9.64&     0.55& 9.55&     \textbf{0.51}  & \textbf{9.51}&\\

\hline

\end{tabular}
}
\end{table}

\section{Experiments}
\label{section:experiments}

We compare the performance of the ILP with the BIC, the AIC and the KN in a series of simulations with $\lambda=\sigma^2=1$. As suggested in \cite{Nadler:2010}, the confidence level $\alpha=10^{-4}$ was used in the KN. Our performance measure is the probability of the successful recovery of the number of signals,
\[
Pr(\hat{k}=k).
\]
We restrict our attention to candidate values for the true number of signals in the range $k'\in\{0,1,\ldots,\min\{p-1,15\}\}$ in simulations. Each simulation in this section is repeated 200 times.

In our experiments, we define $x(t)$ as
\[ \x(t)=\bSigma^{1/2}\z(t),\]
where $\bSigma=\diag\{\lambda_1,\cdots, \lambda_k,\lambda,\cdots,\lambda\}$ and $\z(t)$ is distributed as multivariate normal with mean $0$ and covariance matrix $\I_p$. Thus, the covariance matrix of $\x(t)$ is given by $\bSigma$.

\subsection{Fixed $p$}

\emph{Simulation 1.} In this simulaiton, we investigate how the accuracy of the ILP changes as the SNR varies.
We set $p=12$, $k=3$, $\gamma=1$, $\lambda_1=\cdots=\lambda_{k-1}=1+2SNR$, $\lambda_k=1+SNR$ and $\lambda=1$. We let $SNR=\delta\sqrt{4(p-k/2+1/2)\log\log n/n}$, where $\delta$ increases from $1$ to $2$.

We compare the ILP with the BIC, the AIC, the modified AIC and the KN.  From Table \ref{tab:1a}, we can see that, when $\delta\geq1.5$, the ILP outperforms other methods in general.

\subsection{Infinite $p$}
Note that the AIC-type method with $\gamma>\varphi(c)$ is called the EAIC-type method in this paper. In simulations, we set $\gamma=1.1\varphi(c)$.

\emph{Simulation 2.} In this simulation, in the case $n,p\to\infty$ with $p<n$, we investigate how the accuracy of the EAIC-type method changes as the SNR varies.
We set $n=500$, $p=200$, $k=10$, $\lambda_1=\cdots=\lambda_{k-1}=1+2SNR$, $\lambda_k=1+SNR$ and $\lambda=1$. Note that when $c=p/n=0.4$, $\gamma=1.1\varphi(c)=0.94$ and $\sqrt{c}=0.632$. We let $\delta$ increase from $0.5$ to $2.5$. We also note that when $\delta=0.5$, (\ref{eq:consistency1}) does not hold, while when $1\leq\delta\leq2.5$, (\ref{eq:consistency1}) holds. It can be seen from Table \ref{tab:5a} that, when
$\delta=0.5$, the success rate of the EAIC-type method is low. On the other hand, when $\delta\geq 1$, the success rate is high. This simulation verifies that Theorem \ref{theorem:4} is correct, i.e., if (\ref{eq:consistency1}) and (\ref{eq:consistency2}) hold, the EAIC-type method is consistent. It can also be seen from Table \ref{tab:5a} that, in the case $n,p\to\infty$ with $p<n$, the EAIC-type method is  better than the KN and the BCF , especially for $\delta\leq1$.

Note that when $\delta=1$, (\ref{eq:consistency1}) holds while (\ref{eq:consistency3}) does not hold. Thus, it not surprising that in this case, the EAIC-type method is better than the BCF. On the other hand, although (\ref{eq:consistency1}) holds, $\psi(\lambda_k)-1-\log \psi(\lambda_k)-2\gamma c=0.017$ is close to zero. As a result, when
$\delta=1$, the success rate of the EAIC-type method is not high.

\emph{Simulation 3.} In this simulation, in the case $n,p\to\infty$ with $p>n$, we investigate how the accuracy of the EAIC-type method changes as the SNR varies.
We set $n=200$, $p=500$, $k=10$, $\lambda_1=\cdots=\lambda_{k-1}=1+2SNR$, $\lambda_k=1+SNR$ and $\lambda=1$. Note that when $c=p/n=2.5$, $\gamma=1.1\varphi(c)=0.83$ and $\sqrt{c}=1.581$. We let $\delta$ increase from $1.5$ to $4.5$. We also note that when $\delta=1.5, 2.5$, (\ref{eq:consistency1}) does not hold, while when $2.68\leq\delta\leq4.5$, (\ref{eq:consistency1}) holds. It can be seen from Table \ref{tab:5b} that, when
$\delta=1.5, 2.5$, the success rate of the EAIC-type method is low. On the other hand, when $\delta\geq 3.5$, the success rate is high. This simulation verifies that Theorem \ref{theorem:4} is correct, i.e., if (\ref{eq:consistency1}) and (\ref{eq:consistency2}) hold, the EAIC-type method is consistent. It can also be seen from Table \ref{tab:5b} that, in the case $n,p\to\infty$ with $p>n$, the EAIC-type method is  better than the KN and the BCF, especially for $\delta\leq3.5$.

Note that when $\delta=2.68$, (\ref{eq:consistency1}) holds while (\ref{eq:consistency4}) does not hold. Thus, it not surprising that in this case, the EAIC-type method is better than the BCF. On the other hand, although (\ref{eq:consistency1}) holds, $\psi(\lambda_k)-1-\log \psi(\lambda_k)-2\gamma c=0.0085$ is close to zero. As a result, when
$\delta=2.68$, the success rate of the EAIC-type method is not high.

\emph{Simulation 4.} In this simulation, in the case $n,p\to\infty$ with $p=n$, we investigate how the accuracy of the EAIC-type method changes as the SNR varies.
We set $n=200$, $p=200$, $k=10$, $\lambda_1=\cdots=\lambda_{k-1}=1+2SNR$, $\lambda_k=1+SNR$ and $\lambda=1$. Note that when $c=p/n=1$, $\gamma=1.1\varphi(c)=0.89$ and $\sqrt{c}=1$. We let $\delta$ increase from $1$ to $3$. We also note that when $\delta=1,1.5$, (\ref{eq:consistency1}) does not hold, while when $2\leq\delta\leq3$, (\ref{eq:consistency1}) holds. It can be seen from Table \ref{tab:5c} that, when
$\delta=1,1.5$, the success rate of the EAIC-type method is low. On the other hand, when $\delta\geq 2$, the success rate is high. This simulation verifies that Theorem \ref{theorem:4} is correct, i.e., if (\ref{eq:consistency1}) and (\ref{eq:consistency2}) hold, the EAIC-type method is consistent. It can also be seen from Table \ref{tab:5c} that, in the case $n,p\to\infty$ with $p=n$, the EAIC-type method is better than the KN and the BCF, especially for $\delta\leq2.5$.

Note that when $\delta=2$, (\ref{eq:consistency1}) holds while (\ref{eq:consistency3}) does not hold. Thus, it not surprising that in this case, the EAIC-type method significantly outperforms  the BCF.

\emph{Simulation 5.} In this simulation, in the case $n,p\to\infty$, at low SNR, i.e., $SNR=2\sqrt{p/n}$, we further compare the EAIC-type method with the BCF.
We set $k=10$, $\lambda_1=\cdots=\lambda_{k-1}=1+2SNR$, $\lambda_k=1+SNR$ and $\lambda=1$. It can be seen from Tables \ref{tab:5d} and  \ref{tab:5e} that, when
$c=p/n$ is less than or close to one, the EAIC-type method outperforms the BCF. On the other hand, when $c=p/n$ is larger than one, the EAIC-type method is still comparable to the BCF.

\section{Discussion}
\label{section:discussion}
In this paper, we have demonstrated that any penalty term of the form $k'(p-k'/2+1/2)C_n$ may lead to an asymptotically consistent estimator under the condition that $C_n\to\infty$ and $C_n/n\to0$. Compared with the condition in \cite{Zhao:Krishnaiah:Bai:1986}, i.e., $C_n/\log\log n\to\infty$ and $C_n/n\to0$, this condition has been significantly weakened.  We have also extended our results to the case $n,p\to\infty$, with $p/n\to c>0$. In this case, for increasing $k$, we have investigated the limiting law for the leading eigenvalues of the sample covariance matrix $\S_n$ under the condition that $\lambda_k>1+\sqrt{c}$. This includes the case $\lambda_k\rightarrow\infty$. At low SNR, since the AIC tends to underestimate the number of signals $k$,  the AIC should be re-defined in this case. As a natural extension of the AIC for fixed $p$, we have proposed the extended AIC (EAIC), i.e., the AIC-type method with tuning parameter $\gamma=\varphi(c)=1/2+\sqrt{1/c}-\log(1+\sqrt{c})/c$, and demonstrated that the EAIC-type method, i.e., the AIC-type method with tuning parameter $\gamma>\varphi(c)$, can select the number of signals $k$ consistently. In simulations, we set $\gamma=1.1\varphi(c)$ which gives good performance and outperforms the KN proposed in \cite{Kritchman:Nadler:2008} and the BCF proposed in \cite{Bai:Choi:Fujikoshi:2018}. That is, the EAIC-type method is essentially tuning-free.

We have noted that in the case $n,p\to\infty$ with $p/n\to c>0$, if (\ref{eq:consistency1}) or (\ref{eq:consistency2}) does not hold, $\gamma=1.1\varphi(c)$ tends to underestimate the number of of principal components $k$. As a result, $0<\gamma<1.1\varphi(c)$ may be a better choice. In this case, we may use the cross-validation method to choose the tuning parameter $\gamma$, which will be explored in  future work.

\section{Appendix}\label{section:proof}
This section presents auxiliary lemmas used in this paper and proofs for main results stated in this paper.
\subsection{Auxiliary Lemmas}

\renewcommand{\thelemma}{A.\arabic{lemma}}
\setcounter{lemma}{0}
\begin{lemma}[Theorem 6.3.5 in \cite{Horn:Johnson:2013}]
\label{lemma:perturbation}
Let $\A,\E\in\M_n$, assume that $\A,\A+\E$ are both normal, let $\lambda_1,\ldots,\lambda_n$ be the eigenvalues of $\A$ in some given order, and let $\hat{\lambda}_1,\ldots,\hat{\lambda}_n$ be the eigenvalues of $\A+\E$ in some given order. There is a permutation $\sigma(\cdot)$ of the integers $1,\ldots,n$ such that
\[
\sum_{l=1}^k(\hat{\lambda}_{\sigma(i)}-\lambda_i)^2\leq ||\E||_2^2=\tr(\E^*\E).
\]
\end{lemma}

\begin{lemma}[Lemma 1 in \cite{Cai:Ren:Zhou:2016}]
\label{lemma:spectral}
Suppose $\Y=(\Y_1,\cdots, \Y_p)'$ is sub-Gaussian with constant $\rho>0$ and with mean $0$ and covariance matrix $\bSigma$. Let $\Y^{(1)},\cdots, \Y^{(n)}$ be $n$ independent copies of $Y$. Then there exist some universal constant $C>0$ and some constant $\rho_1$ depending $\rho$, such that the sample covariance matrix of $\Y^{(1)},\cdots, \Y^{(n)}$, $\S$, satisfies
\[
P(\|\S-\bSigma\|>t)\leq 2\exp(-nt^2\rho_1+Cp),
\]
for all $0<t<\rho_1$. Here $\|\cdot\|$ is the spectral norm.
\end{lemma}
For more work on this topic, the reader is referred to \cite{Baik:Arous:Peche:2005}, \cite{Baik:Silverstein:2006}, \cite{Paul:2007} and \cite{Bai:Yao:2008}.

\begin{lemma}[Weyl's inequality in \cite{Weyl:1912}]
\label{lemma:weyl}
Let $\A$ and $\B$ be $p\times p$ Hermitian matrices with eigenvalues ordered as $\lambda_1\geq \cdots\geq\lambda_p$ and $\mu_1\geq\cdots\geq\mu_p$, respectively. Then
\[
\sup_{1\leq j\leq p}\mid\mu_j-\lambda_j\mid\leq\|\A-\B\|,
\]
where $\|\cdot\|$ is the spectral norm.
\end{lemma}

\subsection{Proofs for Main Results}

\begin{proof} [\bf Proof of Lemma \ref{lemma:3}]

By Lemmas \ref{lemma:spectral} and \ref{lemma:weyl}, for $i=1,2,\cdots, p$,
\[
\mid\lambda_i-d_i\mid\leq \max_{1\leq i\leq p}\mid\phi_i\mid=||\S_n-\bSigma||=O_p(\sqrt{p/n}).
\]
Thus, for $i=1,2,\cdots, p$, we have
\[
(\lambda_i-d_i)^2=O_p(p/n)=O_p(1/n).
\]
For fixed $p$, by Lemmas \ref{lemma:spectral} and \ref{lemma:weyl},
\begin{eqnarray*}(p-k)(\lambda-\hat{\lambda})^2&=&(p-k)(\lambda-\frac{1}{p-k}\sum_{i=k+1}^pd_i)^2\\
&=&(\sum_{i=k+1}^p(\lambda-d_i))^2/(p-k)\\
&\leq&p\max_{1\leq i\leq p}(d_i-\lambda_i)^2\\
&\leq&p\max_{1\leq i\leq p}\mid\phi_i\mid^2\\
&=&p||\S_n-\bSigma||^2\\
&=&O_p(p^2/n)
\end{eqnarray*}
\end{proof}

\begin{proof}[\bf Proof of Lemma \ref{lemma:4}]

Noting that $(p-k)(\lambda-\hat{\lambda})^2=O_p(1/n)$, by Taylor's expansion, we get
\begin{eqnarray*}\log L_k&=&-\frac{1}{2}n(\sum_{i=1}^k\log d_i+(p-k)\log\hat{\lambda})\\
&=&-\frac{1}{2}n\sum_{i=1}^k\log d_i-\frac{1}{2}n(p-k)\log((\frac{1}{p-k}\sum_{i=k+1}^pd_i-\lambda)+\lambda)\\
&=&-\frac{1}{2}n\sum_{i=1}^k\log d_i-\frac{1}{2}n(p-k)\log((\frac{1}{p-k}\sum_{i=k+1}^pd_i-1)+1)\\
&=&-\frac{1}{2}n\sum_{i=1}^k\log d_i-\frac{1}{2}n(p-k)((\frac{1}{p-k}\sum_{i=k+1}^pd_i-1)\\
&&-\frac{1}{2}(\frac{1}{p-k}\sum_{i=k+1}^pd_i-1)^2(1+o(1)))\\
&=&-\frac{1}{2}n\sum_{i=1}^k\log d_i-\frac{1}{2}n\sum_{i=k+1}^pd_i+\frac{1}{2}n(p-k)+O_p(1)\\
&=&-\frac{1}{2}n\sum_{i=1}^k\log d_i-\frac{1}{2}n\sum_{i=k+1}^p(d_i-1)+O_p(1).
\end{eqnarray*}
Noting that $C_n\rightarrow \infty$, (\ref{eq:penalized:ILP22}) holds.
\end{proof}

\begin{proof}[\bf Proof of Theorem \ref{theorem:2}]

Let $\log \tilde{L}_k=-\frac{1}{2}n\sum_{i=1}^k\log d_i-\frac{1}{2}n\sum_{i=k+1}^p(d_i-1)$.

Suppose $k'<k$. We have
\[
\tilde{\ell}(k)-\tilde{\ell}(k')=\log \tilde{L}_k-\log \tilde{L}_{k'}-\gamma(k-k')(p-(k+k')/2+1/2)C_n.
\]

By Lemmas \ref{lemma:spectral} and \ref{lemma:weyl}, for $i=1,2,\cdots, p$,
\[
\mid\lambda_i-d_i\mid\leq \max_{1\leq i\leq p}\mid\phi_i\mid=||\S_n-\bSigma||=O_p(\sqrt{1/n}).
\]

Since $\lambda_k>1$, for large $n$,
\[
\lambda_k-1-\log \lambda_k>\epsilon,
\]
where $\epsilon>0$ is a sufficiently small constant.

Since
\[
(d_i-1-\log d_i)-(\lambda_i-1-\log \lambda_i)\stackrel{P}{\longrightarrow}0,
\]
for large $n$, we have
\begin{eqnarray*}P\{\tilde{\ell}(k)>\tilde{\ell}(k')\}&=&P\{\log \tilde{L}_k-\log \tilde{L}_{k'}>\gamma(k-k')(p-(k+k')/2+1/2)C_n\}\\
&=&P\{\frac{1}{2}n\sum_{i=k'+1}^k(d_i-1-\log d_i)>\gamma(k-k')(p-(k+k')/2+1/2)C_n\}\\
&\geq&P\{\frac{1}{2}n(k-k')(\lambda_k-1-\log \lambda_k)+o_p(n)\\
&&>\gamma(k-k')(p-(k+k')/2+1/2)C_n\}\\
&\geq&P\{\frac{1}{2}n(\lambda_k-1-\log \lambda_k)+o_p(n)>\gamma(p-k/2+1/2)C_n\}\\
&\rightarrow&1.
\end{eqnarray*}

Suppose $k'>k$. Since $d_i-1=O_p(\sqrt{1/n})$ for $i>k$, by Taylor's expansion, we get
\begin{eqnarray*}\log \tilde{L}_k-\log \tilde{L}_{k'}&=&\frac{1}{2}n\sum_{i=k+1}^{k'}(\log d_i+1-d_i)\\
&=&\frac{1}{2}n\sum_{i=k+1}^{k'}(\log (1+(d_i-1))+1-d_i)\\
&=&-\frac{1}{4}n\sum_{i=k+1}^{k'}(d_i-1)^2(1+o(1))\\
&=&-O_p(1).
\end{eqnarray*}
Thus,
\begin{eqnarray*}P\{\tilde{\ell}(k)>\tilde{\ell}(k')\}&=&P\{\log \tilde{L}_k-\log \tilde{L}_{k'}>\gamma(k-k')(p-(k+k')/2+1/2)C_n\}\\
&>&P\{-O_p(1)>\gamma(k-k')(p-(k+k')/2+1/2)C_n\}\\
&\rightarrow&1.
\end{eqnarray*}
\end{proof}

\begin{proof}[\bf Proof of Corollary \ref{lemma:large deviation1}]
 Let $\L=\diag\{l_1,\cdots,l_n\}$ and
\begin{eqnarray*}
\A_n=\Q\L\Q^\top,
\end{eqnarray*}
where $\Q$ is orthogonal and its column $\Q_j$ is the eigenvector of $\L$  with eigenvalues $l_j$.
Without loss of generality, we assume that $E(\x_1(t))=\0$. Then,

Let
\[
\Z=\bSigma_{11}^{-\frac{1}{2}}\X_1^\top\Q.
\]
For $1\leq s\leq n$, the $s$-th column of $\Z$,
\[
\Z_s=(\bSigma_{11}^{-\frac{1}{2}}\X_1^\top\Q)_s\sim N(\0, \I),
\]
and are independent and identically distributed.

Noting that
\begin{eqnarray*}
(\frac{1}{n}\bSigma_{11}^{-\frac{1}{2}}\X_1^\top(\I+\A_n)X_1\bSigma_{11}^{-\frac{1}{2}})_{ij}&=&(\frac{1}{n}(\bSigma_{11}^{-\frac{1}{2}}\X_1^\top\Q)(\I+\L)(\Q^\top\X_1\bSigma_{11}^{-\frac{1}{2}}))_{ij}\\
&=&(\frac{1}{n}\Z(\I+\L)\Z^\top)_{ij}\\
&=&\frac{1}{n}\sum_{s=1}^n(1+l_s)z_{is}z_{js},
\end{eqnarray*}
we have
\[
\frac{1}{n}\v^\top\bSigma_{11}^{-\frac{1}{2}}\X_1^\top(\I+\A_n)\X_1\bSigma_{11}^{-\frac{1}{2}}\v=\frac{1}{n}\sum_{s=1}^n(1+l_s)(\sum_{i,j=1}^kv_iv_jz_{is}z_{js}),
\]
where $\v^\top\v=1$. Hence,
\begin{eqnarray*}
E(\frac{1}{n}\bSigma_{11}^{-\frac{1}{2}}\X_1^\top(\I+\A_n)\X_1\bSigma_{11}^{-\frac{1}{2}})_{ij}&=&\frac{1}{n}\sum_{s=1}^n(1+l_s)E(z_{is}z_{js})\\
&=&\frac{1}{n}\sum_{s=1}^n(1+l_s)I_{ij}.
\end{eqnarray*}
That is,
\begin{eqnarray*}
E(\frac{1}{n}\bSigma_{11}^{-\frac{1}{2}}\X_1^\top(\I+\A_n)\X_1\bSigma_{11}^{-\frac{1}{2}})&=&\frac{1}{n}\sum_{s=1}^n(1+l_s)\I=\frac{1}{n}\tr(\I+\A_n)\I.
\end{eqnarray*}
Noting that,
\begin{eqnarray*}
&&\frac{1}{n}\v^\top(\bSigma_{11}^{-\frac{1}{2}}\X_1^\top(I+A_n)\X_1\bSigma_{11}^{-\frac{1}{2}}-\sum_{s=1}^n(1+l_s)\I)\v\\\nonumber
&=&\frac{1}{n}\sum_{s=1}^n(1+l_s)(\sum_{i,j=1}^kv_iv_jz_{is}z_{js})-\frac{1}{n}\sum_{s=1}^n(1+l_s)\v^\top\I\v\\\nonumber
&=&\frac{1}{n}\sum_{s=1}^n(1+l_s)(\sum_{i,j=1}^kv_iv_jz_{is}z_{js})-\frac{1}{n}\sum_{s=1}^n(1+l_s),
\end{eqnarray*}
we have
\begin{eqnarray*}
&&P(\frac{1}{n}\v^\top(\bSigma_{11}^{-\frac{1}{2}}\X_1^\top(\I+\A_n)\X_1\bSigma_{11}^{-\frac{1}{2}}-\tr(\I+\A_n)\I)\v>t)\\\nonumber
&=&P(\v^\top(\bSigma_{11}^{-\frac{1}{2}}\X_1^\top(\I+\A_n)\X_1\bSigma_{11}^{-\frac{1}{2}}-\sum_{s=1}^n(1+l_s)\I)\v>nt)\\\nonumber
&=&\int_\S\prod_{s=1}^ndF(z_{1s},\dots,z_{ks})\\
&\leq&\int_\S\frac{\exp(a\sum_{s=1}^n(1+l_s)(\sum_{i,j=1}^kv_iv_jz_{is}z_{js})-a\sum_{s=1}^n(1+l_s))}{\exp(ant)}\prod_{s=1}^ndF(z_{1s},\dots,z_{ks})\\\nonumber
&\leq&\int_\S\frac{\exp(a\sum_{s=1}^n(1+l_s)(\sum_{i,j=1}^kv_iv_jz_{is}z_{js})-a\sum_{s=1}^n(1+l_s))}{\exp(ant)}\prod_{s=1}^ndF(z_{1s},\dots,z_{ks})\\\nonumber
&=&\exp(-ant-a\sum_{s=1}^n(1+l_s))\prod_{s=1}^nE(\exp(a(1+l_s)\sum_{i,j=1}^kv_iv_jz_{is}z_{js}))\\
&=&\exp(-ant-a\sum_{s=1}^n(1+l_s))\prod_{s=1}^nE(\exp(a(1+l_s)W_s)),
\end{eqnarray*}
where \[\S=\{\sum_{s=1}^n(1+l_s)(\sum_{i,j=1}^kv_iv_jz_{is}z_{js})-\sum_{s=1}^n(1+l_s)>nt\},\]
\[W_s=\sum_{i,j=1}^kv_iv_jz_{is}z_{js}=(\sum_{i=1}^kv_iz_{is})(\sum_{j=1}^kv_jz_{js})=(\sum_{i=1}^kv_iz_{is})^2\sim\chi^2(1).\]

\begin{eqnarray*}
E(\exp(a(1+l_s)W_s))&=&\sum_{i=0}^\infty\int\frac{(a(1+l_s)w)^i}{i!}dF(w)\\
&\leq&1+a(1+l_s)+\sum_{i=2}^\infty\int\frac{(a(1+l_s)w)^i}{i!}dF(w)\\
&\leq&1+a(1+l_s)+\sum_{i=2}^\infty\frac{(a(1+l_s))^iE(w^i)}{i!}\\
&\leq&1+a(1+l_s)+\sum_{i=2}^\infty\frac{(2a(1+l_s))^i}{i!}\frac{\Gamma(\frac{1}{2}+i)}{\Gamma(\frac{1}{2})}\\
&\leq&1+a(1+l_s)+\sum_{i=2}^\infty(2a(1+l_s))^i\\
&\leq&1+a(1+l_s)+4\rho_1a^2(1+l_s)^2,
\end{eqnarray*}
where $2a(1+l_1)<1$, $\rho_1\geq\frac{1}{1-2a(1+l_1)}$.

Hence, for any $0<a<\frac{1}{2(1+l_1)}$, we have
\begin{eqnarray*}
\prod_{s=1}^nE(\exp(a(1+l_s)W_s))&\leq&e^{\sum_{s=1}^n\log(1+a(1+l_s)+4\rho_1a^2(1+l_s)^2)}\\
&\leq&e^{a\sum_{s=1}^n(1+l_s)+4\rho_1a^2\sum_{s=1}^n(1+l_s)^2}.
\end{eqnarray*}

That is, for any $0<a<\frac{1}{2(1+l_1)}$, we have
\begin{eqnarray*}
&&P(\frac{1}{n}\v^\top(\bSigma_{11}^{-\frac{1}{2}}X_1^\top(\I+\A_n)\X_1\bSigma_{11}^{-\frac{1}{2}}-\tr(\I+\A_n)\I)\v>t)\\\nonumber
&\leq&\exp(-ant+4\rho_1a^2\sum_{s=1}^n(1+l_s)^2).
\end{eqnarray*}

Let $f(a)=-ant+4\rho_1a^2\sum_{s=1}^n(1+l_s)^2$.
Setting the derivative $f'(a)$ to zero gives $a=\frac{nt}{8\rho_1\sum_{s=1}^n(1+l_s)^2}$.
Thus,
\begin{eqnarray*}
&&P(\frac{1}{n}\v^\top(\bSigma_{11}^{-\frac{1}{2}}\X_1^\top(\I+\A_n)\X_1\bSigma_{11}^{-\frac{1}{2}}-\tr(\I+\A_n)I)\v>t)\\\nonumber
&\leq&\exp(-\frac{nt^2}{16\rho_1\sum_{s=1}^n(1+l_s)^2/n}).
\end{eqnarray*}

Furthermore, according to the proof of Lemma 3 in \cite{Cai:Zhang:Zhou:2010}, we have
\begin{eqnarray*}
&&P(||\frac{1}{n}\bSigma_{11}^{-\frac{1}{2}}\X_1^\top(\I+\A_n)\X_1\bSigma_{11}^{-\frac{1}{2}}-\frac{1}{n}\tr(\I+\A_n)\I||>t)\\\nonumber
&\leq&P(\sup_{1\leq j\leq5^k}\frac{1}{n}|\v_j^\top(\bSigma_{11}^{-\frac{1}{2}}\X_1^\top(\I+\A_n)\X_1\bSigma_{11}^{-\frac{1}{2}}-\sum_{s=1}^n(1+l_s)\I)\v_j|>\frac{t}{4})\\\nonumber
&\leq&\sum_{j=1}^{5^k}2P(\frac{1}{n}\v_j^\top(\bSigma_{11}^{-\frac{1}{2}}\X_1^\top(\I+\A_n)\X_1\bSigma_{11}^{-\frac{1}{2}}-\sum_{s=1}^n(1+l_s)\I)\v_j>\frac{t}{4})\\\nonumber
&\leq& 2\times5^k\exp(-\frac{nt^2}{16^2\rho_1\sum_{s=1}^n(1+l_s)^2/n})\\\nonumber
&=&2\exp(Ck-\frac{nt^2}{\rho\sum_{s=1}^n(1+l_s)^2/n}),
\end{eqnarray*}
where $\v_j^\top\v_j=1$, $\rho=16^2\rho_1$.

Noting that $2a(1+l_1)<1$, we have
\[
t<4\rho_1\frac{\sum_{s=1}^n(1+l_s)^2/n}{1+l_1}=\rho\frac{\sum_{s=1}^n(1+l_s)^2/n}{64(1+l_1)}.
\]

\end{proof}

\begin{proof}[\bf Proof of Theorem \ref{theorem:a.s.consistency}]
By definition, each $d_i$ solves the equation
\[
0=\mid d \I-\S_n\mid=\mid d \I-\S_{22}\mid\mid d \I-\bm K_n(d)\mid,
\]
where
\[
\bm K_n(d)=\frac{1}{n}\X_1^\top(\I+\H_n(d))\X_1,
\]
\[
\H_n(d)=\frac{1}{n}\X_2(d \I-\S_{22})^{-1}\X_2^\top.
\]
The $d_i$'s then solve the determinant equation
\[
\mid d \I-\bm K_n(d)\mid=0.
\]
That is, $d_i$'s are solutions of
\[
\left| d\I-\frac{1}{n}\X_1^\top(\I+\H_n(d))\X_1\right|=0.
\]
Or equivalently,
\begin{equation}
\label{thm1:eq1}
\left| \bSigma_{11}^{-1}d\I-\frac{1}{n}\bSigma_{11}^{-\frac{1}{2}}\X_1^\top(\I+\H_n(d))\X_1\bSigma_{11}^{-\frac{1}{2}}\right|=0
\end{equation}
Next, we demonstrate that the above determinant \eqref{thm1:eq1} has $k$ solutions in  the interval $((1+\sqrt{c})^2, \infty)$. For $d>(1+\sqrt{c})^2$,
by the definition of $F_c(x)$, it is easy to see that $m_1(d)=O(1)$ and $m_2(d)=O(1)$.

Let $\beta_j$, $j=1,\dots, p-k$ be the eigenvalues of $\S_{22}=\frac{1}{n}\X_2^\top\X_2$. Then we have
\begin{equation}\label{thm1:eq2}
\begin{aligned}
\frac{1}{n}\tr(\H_n(d))&=\frac{1}{n}\tr(\frac{1}{n}\X_2(d \I-\S_{22})^{-1}\X_2^\top)=\frac{1}{n}\tr((d \I-\S_{22})^{-1}\S_{22})\\
&=\frac{1}{n}\sum_{j=1}^{p-k}\frac{\beta_j}{d-\beta_j}=\frac{p-k}{n}\frac{1}{p-k}\sum_{j=1}^{p-k}\frac{\beta_j}{d-\beta_j}\\
&=\frac{p-k}{n}\int\frac{x}{d-x}dF_n(x)=\frac{p-k}{n}\int\frac{x}{d-x}dF_c(x)(1+O_p(\frac{k}{n}))\\
&=(c+(\frac{p}{n}-c)-\frac{k}{n})m_1(d)(1+O_p(\frac{k}{n}))\\
&=cm_1(d)(1+O_p(\sqrt{\frac{k}{n}})).
\end{aligned}
\end{equation}

Moreover, we have
\begin{equation}\label{thm1:eq3}
\begin{aligned}
\frac{1}{n}\tr(\H_n^2(d))&=\frac{1}{n}\tr((d \I-\S_{22})^{-1}\S_{22}(d \I-\S_{22})^{-1}\S_{22})=\frac{1}{n}\sum_{j=1}^{p-k}\frac{\beta_j^2}{(d-\beta_j)^2}\\
&=\frac{p-k}{n}\frac{1}{p-k}\sum_{j=1}^{p-k}\frac{\beta_j^2}{(d-\beta_j)^2}=\frac{p-k}{n}\int\frac{x^2}{(d-x)^2}dF_n(x)\\
&=\frac{p-k}{n}\int\frac{x^2}{(d-x)^2}dF_c(x)(1+O_p(\frac{k}{n}))\\
&=(c+(\frac{p}{n}-c)-\frac{k}{n})m_2(d)(1+O_p(\frac{k}{n}))\\
&=cm_2(d)(1+O_p(\sqrt{\frac{k}{n}})).
\end{aligned}
\end{equation}

Combing \eqref{thm1:eq2} and \eqref{thm1:eq3}, we have
\[
\frac{1}{n}\sum_{s=1}^n(1+l_s)^2=1+\frac{2}{n}\tr(\H_n(d))+\frac{1}{n}\tr(\H_n^2(d))=1+O_p(1),
\]
where $l_1\geq\cdots\geq l_n$ are the eigenvalues of $\H_n(d)$.

By Corollary \ref{lemma:large deviation1}, we have
\[
||\frac{1}{n}\bSigma_{11}^{-\frac{1}{2}}\X_1^\top(\I+\H_n(d))\X_1\bSigma_{11}^{-\frac{1}{2}}-\frac{1}{n}\tr(\I+\H_n(d))I||=O_p(\sqrt{\frac{k}{n}}).
\]

As a result, we have the following spectral decomposition of
\begin{equation}\label{thm1:eq4}
\begin{aligned}
&\frac{1}{n}\bSigma_{11}^{-\frac{1}{2}}\X_1^\top(\I+\H_n(d))\X_1\bSigma_{11}^{-\frac{1}{2}}-\frac{1}{n}\tr(\I+\H_n(d))\I\\
=&\V(d)\diag(v_1(d),\ldots,v_k(d))\V^\top(d),
\end{aligned}
\end{equation}
where $\V(d)$ is an orthogonal matrix and $\max_{1\leq j\leq k}|v_j(d)|=O_p(\sqrt{\frac{k}{n}})$.

By \eqref{thm1:eq2}, we have
\begin{equation}\label{thm1:eq5}
\frac{1}{n}\tr(\I+\H_n(d))=1+cm_1(d)(1+O_p(\sqrt{\frac{k}{n}})),
\end{equation}
Combing \eqref{thm1:eq4} and \eqref{thm1:eq5}, we have
\begin{equation}\label{thm1:eq6}
\begin{aligned}
&\frac{1}{n}\bSigma_{11}^{-\frac{1}{2}}\X_1^\top(\I+\H_n(d))\X_1\bSigma_{11}^{-\frac{1}{2}}-(1+cm_1(d))\I\\
=&\M(d)\triangleq \V(d)\diag(\tilde{v}_1(d),\ldots,\tilde{v}_k(d))\V^\top(d),
\end{aligned}
\end{equation}
where $\max_{1\leq j\leq k}|\tilde{v}_j(d)|=O_p(\sqrt{\frac{k}{n}})$.

Note that
\begin{eqnarray*}
&&\bSigma_{11}^{-1}d\I-\frac{1}{n}\bSigma_{11}^{-\frac{1}{2}}\X_1^\top(\I+\H_n(d))\X_1\bSigma_{11}^{-\frac{1}{2}}\\\nonumber
&=&\bSigma_{11}^{-1}d\I-(1+cm_1(d))\I-[\frac{1}{n}\bSigma_{11}^{-\frac{1}{2}}\X_1^\top(\I+\H_n(d))\X_1\bSigma_{11}^{-\frac{1}{2}}-(1+cm_1(d))\I]\\
&=&\bSigma_{11}^{-1}d\I-(1+cm_1(d))\I-\M(d).
\end{eqnarray*}
and
\[
\begin{array}{lll}\bSigma_{11}^{-1}=\U\diag(\frac{1}{\lambda_1},\ldots,\frac{1}{\lambda_k})\U^\top.
\end{array}
\]

Multiplying both sides of \eqref{thm1:eq1} by $\U^\top$ from the left and by $\U$ from the right yields
\begin{eqnarray*}
&&|\U^\top(\bSigma_{11}^{-1}d\I-\frac{1}{n}\bSigma_{11}^{-\frac{1}{2}}\X_1^\top(\I+\H_n(d))\X_1\bSigma_{11}^{-\frac{1}{2}})\U|\\
&=&|\diag(\frac{d-(1+cm_1(d))\lambda_1}{\lambda_1},\ldots,\frac{d-(1+cm_1(d))\lambda_k}{\lambda_k})-\U^\top\M(d)\U|\\
&=&0.
\end{eqnarray*}
Let $\hat{e}_{\sigma(i)}$'s be the eigenvalues of the following matrix,
\[
\diag(\frac{d-(1+cm_1(d))\lambda_1}{\lambda_1},\ldots,\frac{d-(1+cm_1(d))\lambda_k}{\lambda_k})-\U^\top\M(d)\U,
\]
where $\sigma(\cdot)$ is a permutation  of the integers $1,\ldots,n$.

Then, $d_i$'s are the solutions of
\begin{equation*}
\prod_{i=1}^k\hat{e}_{\sigma(i)}=0
\end{equation*}

That is, for  any $1\leq i\leq k$, $d_i$ is the solution of
\begin{equation}
\label{thm1:eq7}
\hat{e}_{\sigma(i)}=0.
\end{equation}

Also, let $e_{i}$'s be the eigenvalues of the following matrix,
\[
\diag(\frac{d-(1+cm_1(d))\lambda_1}{\lambda_1},\ldots,\frac{d-(1+cm_1(d))\lambda_k}{\lambda_k}).
\]
The following result shows how close are the eigenvalues of $\hat{e}_{\sigma(i)}$'s to $e_i$'s.

Noting that
\[
\U^\top\M(d)\U=\U^\top\V(d)\diag(\tilde{v}_1(d),\ldots,\tilde{v}_2(d))\V^\top(d)\U,
\]

we have
\begin{equation}
\label{thm1:eq8}
\tr((\U^\top\M(d)\U)^2)=\tr(\diag(\tilde{v}_1^2(d),\ldots,\tilde{v}_k^2(d)))=kO_p(\frac{k}{n})=O_p(\frac{k^2}{n}).
\end{equation}
By Lemma \ref{lemma:perturbation}, we have
\[
\sum_{i=1}^k(\hat{e}_{\sigma(i)}-e_i)^2\leq\tr((\U^\top\M(d)\U)^2).
\]
Hence,
\begin{equation}
\label{thm1:eq9}
\sum_{i=1}^k|\hat{e}_{\sigma(i)}-e_i|\leq \sqrt{k\sum_{i=1}^k(\hat{e}_{\sigma(i)}-e_i)^2}=O_p(\sqrt{\frac{k^3}{n}}).
\end{equation}

Combining \eqref{thm1:eq7} and \eqref{thm1:eq9}, for  any $1\leq i\leq k$, $d_i$ tends to the solution of
\begin{equation*}
\frac{d_i-(1+cm_1(d_i))\lambda_i}{\lambda_i}=0.
\end{equation*}

According to \cite{Bai:Yao:2008}, for  any $1\leq i\leq k$, we have
\[
\frac{d_i}{\psi(\lambda_i)}\stackrel{a.s.}{\longrightarrow}1.
\]
\end{proof}

\begin{proof} [\bf Proof of Lemma \ref{lemma:diagonal}]
By definition, each $d_i$ solves the equation
\[
0=\mid d \I-\S_n\mid=\mid d \I-\S_{22}\mid\mid d \I-\bm K_n(d)\mid,
\]
where
\[
\bm K_n(d)=\frac{1}{n}\X_1^\top(\I+\H_n(d))\X_1,
\]
\[
\H_n(d)=\frac{1}{n}\X_2(d \I-\S_{22})^{-1}\X_2^\top.
\]
The $d_i$'s then solve the determinant equation
\[
\mid d \I-\bm K_n(d)\mid=0.
\]
That is
\[
\left| d_i\I-\frac{1}{n}\X_1^\top(\I+\H_n(d_i))\X_1\right|=0.
\]
Or equivalently,
\begin{equation}
\label{thm2:eq1}
\left| \bSigma_{11}^{-1}d_i\I-\frac{1}{n}\bSigma_{11}^{-\frac{1}{2}}\X_1^\top(\I+\H_n(d_i))\X_1\bSigma_{11}^{-\frac{1}{2}}\right|=0
\end{equation}
Note that
\begin{eqnarray*}
&&\Sigma_{11}^{-1}d_i\I-\frac{1}{n}\bSigma_{11}^{-\frac{1}{2}}\X_1^\top(\I+\H_n(d_i))\X_1\bSigma_{11}^{-\frac{1}{2}}\\\nonumber
&=&\Sigma_{11}^{-1}d_i\I-(1+cm_1(d_i))\I-[\frac{1}{n}\bSigma_{11}^{-\frac{1}{2}}\X_1^\top(\I+\H_n(d_i))\X_1\bSigma_{11}^{-\frac{1}{2}}-(1+cm_1(d_i))\I]\\\nonumber
&=&\bSigma_{11}^{-1}d_i\I-(1+cm_1(d_i))\I-\M(d_i),
\end{eqnarray*}
where $\M(d_i)=\frac{1}{n}\bSigma_{11}^{-\frac{1}{2}}\X_1^\top(\I+\H_n(d_i))\X_1\bSigma_{11}^{-\frac{1}{2}}-(1+cm_1(d_i))\I$.
Similar to the proof of \eqref{thm1:eq6} in Theorem \ref{theorem:a.s.consistency}, we have
\[
\M(d_i)=\V(d_i)\diag(\tilde{v}_1(d_i),\ldots,\tilde{v}_k(d_i))\V^\top(d_i),
\]
where $\V(d_i)$ is an orthogonal matrix and $\max_{1\leq j\leq k}|\tilde{v}_j(d_i)|=O_p(\sqrt{\frac{k}{n}})$.

Note that
 \[
\bSigma_{11}^{-1}=\U\diag(\frac{1}{\lambda_1},\ldots,\frac{1}{\lambda_k})\U^\top.
\]

Multiplying both sides of \eqref{thm2:eq1} by $\U^\top$ from the left and by $\U$ from the right yields
\begin{eqnarray*}
&&|\U^\top(\bSigma_{11}^{-1}d_i\I-\frac{1}{n}\bSigma_{11}^{-\frac{1}{2}}\X_1^\top(\I+\H_n(d_i))\X_1\bSigma_{11}^{-\frac{1}{2}})\U|\\\nonumber
&=&|\diag(\frac{d_i-(1+cm_1(d_i))\lambda_1}{\lambda_1},\ldots,\frac{d_i-(1+cm_1(d_i))\lambda_k}{\lambda_k})-\U^\top\M(d_i)\U|\\
&=&0.
\end{eqnarray*}

Let
\begin{equation}\label{thm2:eq2}
\begin{aligned}
&\B(d_i)=(b_{st})_{k\times k}\\
=&\U^\top\M(d_i)\U\\
=&\U^\top\V(d_i)\diag(\tilde{v}_1(d_i),\ldots,\tilde{v}_k(d_i))\V^\top(d_i)\U\\
\triangleq&\Q(d_i)\diag(\tilde{v}_1(d_i),\ldots,\tilde{v}_k(d_i))\Q^\top(d_i),
\end{aligned}
\end{equation}
where $\Q(d_i)=\U^\top\V(d_i)=(q_{st})_{k\times k}$ is an orthogonal matrix. For any $1\leq s\leq k $, $1\leq t\leq k$, and $s\neq t$, by the definitions of $\A(d_i)$ and $\B(d_i)$,   we have $a_{st}=b_{st}$.

Similar to the proof of \eqref{thm1:eq8} in Theorem \ref{theorem:a.s.consistency}, we have
\begin{equation}\label{thm2:eq3}
\begin{aligned}
\sum_{s=1}^k\sum_{t=1}^kb_{st}^2=\tr(\B^2(d_i))=O_p(\frac{k^2}{n}).
\end{aligned}
\end{equation}

We also have
\begin{equation}\label{thm2:eq4}
\begin{aligned}
&\sum_{t=1}^kb_{it}^2=\sum_{t=1}^k(\sum_{l=1}^k\tilde{v}_l(d_i)q_{il}q_{tl})^2\\
=&\sum_{t=1}^k\sum_{l=1}^k\tilde{v}_l^2(d_i)q_{il}^2q_{tl}^2+\sum_{t=1}^k\sum_{l=1}^k\sum_{s\neq l}^k\tilde{v}_l(d_i)\tilde{v}_s(d_i)q_{il}q_{tl}q_{is}q_{ts}\\
=&\sum_{t=1}^k\sum_{l=1}^k\tilde{v}_l^2(d_i)q_{il}^2q_{tl}^2+\sum_{l=1}^k\sum_{s\neq l}^k\tilde{v}_l(d_i)\tilde{v}_s(d_i)q_{il}q_{is}\sum_{t=1}^kq_{tl}q_{ts}\\
=&\sum_{t=1}^k\sum_{l=1}^k\tilde{v}_l^2(d_i)q_{il}^2q_{tl}^2=\sum_{l=1}^k\tilde{v}_l^2(d_i)q_{il}^2\sum_{t=1}^kq_{tl}^2\\
=&\sum_{l=1}^k\tilde{v}_l^2(d_i)q_{il}^2\leq\max_{1\leq l\leq k}\tilde{v}_l^2(d_i)\sum_{l=1}^kq_{il}^2\\
=&O_p(\frac{k}{n}).
\end{aligned}
\end{equation}

Note that
\begin{equation}
\label{thm2:eq5}
\det(\A(d_i))=\sum_{t=1}^k(-1)^{i+t}a_{it}H_{it}=0,
\end{equation}
where the $(k-1)$-th order determinant $H_{it}$ is the cofactor of $a_{it}$, obtained by deleting the $i$-th row and the $t$-th column of the $k$-th order determinant $\det(\A(d_i))$.

Let
\[
H_{ii}=\prod_{j\neq i}\hat{e}_j'\triangleq\prod_{j\neq i}(e_j+\epsilon_j'),
\]
where $\hat{e}_j'(j\neq i)$ are eigenvalues of corresponding matrix $\H_{ii}$.
Noting that $\H_{ii}$ is a symmetric matrix, similar to the proof of \eqref{thm1:eq9} in Theorem \ref{theorem:a.s.consistency}, we have
\[
\sum_{j\neq i}^k|\epsilon_j'|=O_p(\sqrt{\frac{k^3}{n}}).
\]

Noting that
\begin{eqnarray*}|\sum_{t\neq i}\frac{1}{e_t}\epsilon_t'|&\leq&\sum_{t\neq i}\frac{1}{|e_t|}|\epsilon_t'|\\
&\leq&\frac{1}{\min_{t\neq i}|e_t|}\sum_{t\neq i}|\epsilon_t'|\\
&\leq&\frac{1}{\min_{t\neq i}|e_t|}O_p(\sqrt{\frac{k^3}{n}})\\
&=&O_p(\sqrt{\frac{k^3}{n}}),
\end{eqnarray*}
and
\begin{eqnarray*}|\sum_{t\neq i}\sum_{s\neq i,1\leq s<t}\frac{1}{e_te_s}\epsilon_t'\epsilon_s'|&\leq&\sum_{t\neq i}\sum_{s\neq i,1\leq s<t}\frac{1}{|e_te_s|}|\epsilon_t'\epsilon_s'|\\
&\leq&\frac{1}{\min_{t\neq i,s\neq i,1\leq s<t}|e_te_s|}\sum_{t\neq i}\sum_{s\neq i,1\leq s<t}|\epsilon_t'\epsilon_s'|\\
&\leq&\frac{1}{\min_{t\neq i,s\neq i,1\leq s<t}|e_te_s|}(\sum_{t\neq i}|\epsilon_t'|)^2\\
&=&\frac{1}{\min_{t\neq i,s\neq i,1\leq s<t}|e_te_s|}O_p(\frac{k^3}{n})\\
&=&O_p(\frac{k^3}{n}),
\end{eqnarray*}
we have
\begin{equation}\label{thm2:eq6}
\begin{aligned}
H_{ii}&=\prod_{j\neq i}(e_j+\epsilon_j')\\
&=\prod_{j\neq i}e_j+\prod_{j\neq i}e_j\sum_{t\neq i}\frac{1}{e_t}\epsilon_t'+\prod_{j\neq i}e_j\sum_{t\neq i}\sum_{s\neq i,1\leq s<t}\frac{1}{e_te_s}\epsilon_t'\epsilon_s'+\cdots\\
&=\prod_{j\neq i}e_j+O_p(\sqrt{\frac{k^3}{n}})\prod_{j\neq i}e_j+O_p(\frac{k^3}{n})\prod_{j\neq i}e_j+\cdots\\
&=(1+O_p(\sqrt{\frac{k^3}{n}}))\prod_{j\neq i}e_j.
\end{aligned}
\end{equation}

For $t\neq i$,
 \[
H_{it}=\prod_{j\neq i, t}^k\hat{e}_j''\hat{e}_{t}'',
\]
where $\hat{e}_j''(j\neq i, t)$ and $\hat{e}_{t}''$ are eigenvalues of corresponding matrix $\H_{it}$.

By Gerschgorin's circle theorem, we have
\[
|H_{it}|=\prod_{j\neq i, t}^k|\hat{e}_j'||\hat{e}_{t}''|\leq\prod_{j\neq i, t}^k(|e_j|+\sum_{r=1}^k|b_{jr}|)|\hat{e}_{t}''|\leq\prod_{j\neq i, t}^k(|e_j|+\sum_{r=1}^k|b_{jr}|)|\sum_{r\neq t}^k|a_{tr}|.
\]

Noting that
\begin{eqnarray*}\sum_{s\neq i,t}\frac{1}{|e_s|}\sum_{r=1}^k|b_{sr}|&\leq&\frac{1}{\min_{s\neq i,t}|e_s|}\sum_{s\neq i,t}\sum_{r=1}^k|b_{sr}|\\
&\leq&\frac{1}{\min_{s\neq i,t}|e_s|}\sqrt{k^2\sum_{s\neq i,t}\sum_{r=1}^kb_{sr}^2}\\
&=&O_p(\sqrt{\frac{k^4}{n}}),
\end{eqnarray*}
and
\begin{eqnarray*}&&\sum_{s\neq i,t}\sum_{u\neq i,t,1\leq u<s}\frac{1}{|e_se_u|}\sum_{r=1}^k|b_{sr}|\sum_{r=1}^k|b_{ur}|\\
&\leq&\frac{1}{\min_{s\neq i,t,u\neq i,t,1\leq u<s}|e_se_u|}\sum_{s\neq i,t}\sum_{r=1}^k|b_{sr}|\sum_{u\neq i,t,1\leq u<s}\sum_{r=1}^k|b_{ur}|\\
&=&\frac{1}{\min_{s\neq i,t,u\neq i,t,1\leq u<s}|e_se_u|}O_p(\frac{k^4}{n})\\
&=&O_p(\frac{k^4}{n}),
\end{eqnarray*}
we have
\begin{eqnarray*}&&\prod_{j\neq i, t}^k(|e_j|+\sum_{r=1}^k|b_{jr}|)\\
&=&\prod_{j\neq i,t}|e_j|+\prod_{j\neq i,t}|e_j|\sum_{s\neq i,t}\frac{1}{|e_s|}\sum_{r=1}^k|b_{sr}|+\prod_{j\neq i,t}|e_j|\sum_{s\neq i,t}\sum_{u\neq i,t,1\leq u<s}\frac{1}{|e_se_u|}\sum_{r=1}^k|b_{sr}|\sum_{r=1}^k|b_{ur}|+\cdots\\
&=&\prod_{j\neq i,t}|e_j|+O_p(\sqrt{\frac{k^4}{n}})\prod_{j\neq i,t}|e_j|+O_p(\frac{k^4}{n})\prod_{j\neq i,t}|e_j|+\cdots\\
&=&(1+O_p(\sqrt{\frac{k^4}{n}}))\prod_{j\neq i,t}|e_j|\\
&\leq&\frac{1+O_p(\sqrt{\frac{k^4}{n}})}{\min_{j\neq i}|e_j|}\prod_{j\neq i}|e_j|\\
\end{eqnarray*}
Hence,
\begin{equation}\label{thm2:eq7}
|H_{it}|\leq\frac{1+O_p(\sqrt{\frac{k^4}{n}})}{\min_{j\neq i}|e_j|}\prod_{j\neq i}|e_j||\hat{e}_{t}''|.
\end{equation}
Combing \eqref{thm2:eq3}, \eqref{thm2:eq4}, \eqref{thm2:eq7}, and noting that for any $1\leq s\leq k $, $1\leq t\leq k$, $s\neq t$, $a_{st}=b_{st}$,
we have
\begin{equation}\label{thm2:eq8}
\begin{aligned}
|\sum_{t\neq i}^ka_{it}H_{it}|&\leq\frac{1+O_p(\sqrt{\frac{k^4}{n}})}{\min_{j\neq i}|e_j|}\prod_{j\neq i}^k|e_j|\sum_{t\neq i}^k|a_{it}\hat{e}_{t}''|\\
&\leq\frac{1+O_p(\sqrt{\frac{k^4}{n}})}{\min_{j\neq i}|e_j|}\prod_{j\neq i}^k|e_j|\sqrt{\sum_{t\neq i}^ka_{it}^2\sum_{t\neq i}^k\hat{e}_{t}''^2}\\
&\leq\frac{1+O_p(\sqrt{\frac{k^4}{n}})}{\min_{j\neq i}|e_j|}\prod_{j\neq i}^k|e_j|\sqrt{\sum_{t\neq i}^ka_{it}^2\sum_{t\neq i}^k(\sum_{r\neq t}^k|a_{tr}|)^2}\\
&\leq\frac{1+O_p(\sqrt{\frac{k^4}{n}})}{\min_{j\neq i}|e_j|}\prod_{j\neq i}^k|e_j|\sqrt{\sum_{t\neq i}^kb_{it}^2\sum_{t\neq i}^k(\sum_{r\neq t}^k|b_{tr}|)^2}\\
&\leq\frac{1+O_p(\sqrt{\frac{k^4}{n}})}{\min_{j\neq i}|e_j|}\prod_{j\neq i}^k|e_j|\sqrt{\sum_{t\neq i}^kb_{it}^2\sum_{t\neq i}^kk\sum_{r\neq t}^kb_{tr}^2}\\
&=\frac{1+O_p(\sqrt{\frac{k^4}{n}})}{\min_{j\neq i}|e_j|}O_p(\sqrt{\frac{k^4}{n^2}})\prod_{j\neq i}^k|e_j|.
\end{aligned}
\end{equation}
Combing \eqref{thm2:eq5},  \eqref{thm2:eq6}, and \eqref{thm2:eq8}, we have
\begin{eqnarray*}|a_{ii}|&=&|\frac{\sum_{t\neq i}^ka_{it}H_{it}}{H_{ii}}|\\
&\leq&\frac{1+O_p(\sqrt{\frac{k^4}{n}})}{1+O_p(\sqrt{\frac{k^3}{n}})}\frac{1}{\min_{j\neq i}|e_j|}O_p(\sqrt{\frac{k^4}{n^2}})\\
&=&\frac{1+O_p(\sqrt{\frac{k^4}{n}})}{\min_{j\neq i}|e_j|}O_p(\sqrt{\frac{k^4}{n^2}})\\
&=&O_p(\sqrt{\frac{k^4}{n^2}}).
\end{eqnarray*}

Thus,  for any $1\leq i\leq k$, the $i$-th diagonal element of $\A$ satisfies
\[
\frac{d_i-(1+cm_1(d_i))\lambda_i}{\lambda_i}-[\U^\top\M(d_i)\U]_{ii}=O_p(\sqrt{\frac{k^4}{n^2}}).
\]

\end{proof}

\begin{proof}[\bf Proof of Theorem \ref{theorem:root-n/k-consistency}]
According to \eqref{thm2:eq2} in Lemma \ref{lemma:diagonal}, we have
\[
[\U^\top\M(d_i)\U]_{ii}=b_{ii}=\sum_{l=1}^k\tilde{v}_l(d_i)q_{il}^2=O_p(\sqrt{\frac{k}{n}}).
\]
By Lemma \ref{lemma:diagonal}, we have
\begin{equation}
\label{thm2:eq9}
\frac{d_i-(1+cm_1(d_i))\lambda_i}{\lambda_i}=O_p(\sqrt{\frac{k}{n}}).
\end{equation}
For  any $1\leq i\leq k$, by Theorem \ref{theorem:a.s.consistency}, we have $d_i>(1+\sqrt{c})^2$. Let $d_i=\alpha_i+\frac{c\alpha_i}{\alpha_i-1}\triangleq\psi(\alpha_i)$.  Then, we get
\[
\alpha_i=\frac{-c+d_i+1+\sqrt{(c-d_i-1)^2-4d_i}}{2},
\]
\[
\frac{\d\alpha_i}{\d d_i}=\frac{1}{2}+\frac{1}{2}\frac{d_i-c-1}{\sqrt{(c-d_i-1)^2-4d_i}}>0.
\]
As a result, we have
$\alpha_i>1+\sqrt{c}$. According to \cite{Bai:Yao:2008}, $m_1(d_i)=m_1(\psi(\alpha_i))=\frac{1}{\alpha_i-1}$. Thus, \eqref{thm2:eq9} becomes
\[
(\frac{\alpha_i}{\lambda_i}-1)(1+\frac{c}{\alpha_i-1})=O_p(\sqrt{\frac{k}{n}}).
\]
That is,
\[
\alpha_i=\lambda_i(1+O_p(\sqrt{\frac{k}{n}})).
\]
Thus, for  any $1\leq i\leq k$, we have
\begin{eqnarray*}\frac{d_i-\psi(\lambda_i)}{\lambda_i}&=&\frac{\alpha_i+\frac{c\alpha_i}{\alpha_i-1}-\lambda_i-\frac{c\lambda_i}{\lambda_i-1}}{\lambda_i}\\
&=&1+O_p(\sqrt{\frac{k}{n}})+\frac{c(1+O_p(\sqrt{\frac{k}{n}}))}{\lambda_i(1+O_p(\sqrt{\frac{k}{n}}))-1}-1-\frac{c}{\lambda_i-1}\\
&=&O_p(\sqrt{\frac{k}{n}})-\frac{cO_p(\sqrt{\frac{k}{n}})}{(\lambda_i(1+O_p(\sqrt{\frac{k}{n}}))-1)(\lambda_i-1)}\\
&=&O_p(\sqrt{\frac{k}{n}}).
\end{eqnarray*}
\end{proof}

\begin{proof}[\bf Proof of Theorem \ref{theorem:asymptotical normal}]
By definition, each $d_i$ solves the equation
\[
0=\mid d \I-\S_n\mid=\mid d \I-\S_{22}\mid\mid d \I-\bm K_n(d)\mid,
\]
where
\[
\bm K_n(d)=\frac{1}{n}\X_1^\top(\I+\H_n(d))\X_1,
\]
\[
\H_n(d)=\frac{1}{n}\X_2(d \I-\S_{22})^{-1}\X_2^\top.
\]
The $d_i$'s then solve the determinant equation
\[
\mid d \I-\bm K_n(d)\mid=0.
\]
That is
\[
\left| d_i\I-\frac{1}{n}\X_1^\top(\I+\H_n(d_i))\X_1\right|=0.
\]

The random form $K_n(d_i)$ can be decomposed as follows
\begin{equation}\label{thm3:eq8}
\begin{aligned}
&\bm K_n(d_i)\frac{1}{n}\X_1^\top(\I+\H_n(d_i))\X_1\\
=&\frac{1}{n}(\X_1^\top(\I+\H_n(d_i))\X_1-\tr(\I+\H_n(d_i))\bSigma_{11})+\frac{1}{n}\tr(\I+\H_n(d_i))\bSigma_{11}\\
=&\frac{1}{\sqrt{n}}R_n(d_i)+\frac{1}{n}\tr(\I+\H_n(d_i))\bSigma_{11},
\end{aligned}
\end{equation}
where
\begin{equation}
\bm R_n(d_i)=\frac{1}{\sqrt{n}}(\X_1^\top(\I+\H_n(d_i))\X_1-\tr(\I+\H_n(d_i))\bSigma_{11}).\nonumber
\end{equation}
Let
\[
\delta_i=\frac{\sqrt{n}(d_i-\psi(\lambda_i))}{\lambda_i}=O_p(\sqrt{k}).
\]

Then the form $d_i \I-\bm K_n(d_i)$ can be decomposed as follows
\begin{equation}
\label{thm3:eq1}
d_i \I-\bm K_n(d_i)=\psi(\lambda_i)\I+\frac{1}{\sqrt{n}}\delta_i\lambda_i\I-\bm K_n(\psi(\lambda_i))-(\bm K_n(d_i)-\bm K_n(\psi(\lambda_i))).
\end{equation}

Furthermore, using $\A^{-1}-\B^{-1}=\A^{-1}(\B-\A)\B^{-1}$ and treating $\A=(\psi(\lambda_i)+\frac{1}{\sqrt{n}}\delta_i\lambda_i)\I-\S_{22}$, $\B=\psi(\lambda_i)\I-\S_{22}$, we have
\begin{equation}\label{thm3:eq2}
\begin{aligned}
&\bSigma_{11}^{-\frac{1}{2}}[\bm K_n(d_i)-\bm K_n(\psi(\lambda_i))]\bSigma_{11}^{-\frac{1}{2}}\\
=&\bSigma_{11}^{-\frac{1}{2}}\frac{1}{n}\X_1^\top\frac{1}{n}\X_2\{[(\psi(\lambda_i)+\frac{1}{\sqrt{n}}\delta_i\lambda_i)\I-\S_{22}]^{-1}\\
&-(\psi(\lambda_i)\I-\S_{22})^{-1}\}\X_2^\top\X_1\bSigma_{11}^{-\frac{1}{2}}\\
=&-\frac{1}{\sqrt{n}}\delta_i\lambda_i\bSigma_{11}^{-\frac{1}{2}}\frac{1}{n}\X_1^\top\frac{1}{n}\X_2\{[(\psi(\lambda_i)+\frac{1}{\sqrt{n}}\delta_i\lambda_i)\I-\S_{22}]^{-1}\\
&[\psi(\lambda_i)\I-\S_{22}]^{-1}\}\X_2^\top\X_1\bSigma_{11}^{-\frac{1}{2}}.
\end{aligned}
\end{equation}
Treating
\begin{equation}\label{thm3:eq3}
\A_n=\lambda_i\frac{1}{n}\X_2[(\psi(\lambda_i)+\frac{1}{\sqrt{n}}\delta_i\lambda_i)\I-\S_{22}]^{-1}(\psi(\lambda_i)\I-\S_{22})^{-1}\X_2^\top,
\end{equation}

we have
\begin{equation}\label{thm3:eq4}
\begin{aligned}
&\frac{1}{n}\tr(\A_n)\\
=&\frac{1}{n}\tr\{\lambda_i\frac{1}{n}\X_2[(\psi(\lambda_i)+\frac{1}{\sqrt{n}}\delta_i\lambda_i)\I-\S_{22}]^{-1}(\psi(\lambda_i)\I-\S_{22})^{-1}\X_2^\top\}\\
=&\lambda_i\frac{1}{n}\tr\{[(\psi(\lambda_i)+\frac{1}{\sqrt{n}}\delta_i\lambda_i)\I-\S_{22}]^{-1}(\psi(\lambda_i)\I-\S_{22})^{-1}\S_{22}\}\\
=&\lambda_i\frac{1}{n}\sum_{j=1}^{p-k}\frac{\beta_j}{(\psi(\lambda_i)+\frac{1}{\sqrt{n}}\delta_i\lambda_i-\beta_j)(\psi(\lambda_i)-\beta_j)}\\
=&\lambda_i\frac{p-k}{n}\frac{1}{p-k}\sum_{j=1}^{p-k}\frac{\beta_j}{(\psi(\lambda_i)+\frac{1}{\sqrt{n}}\delta_i\lambda_i-\beta_j)(\psi(\lambda_i)-\beta_j)}\\
=&\lambda_i\frac{p-k}{n}\frac{1}{p-k}\sum_{j=1}^{p-k}\frac{\beta_j}{(\psi(\lambda_i)-\beta_j)^2}\frac{(\psi(\lambda_i)-\beta_j)}{(\psi(\lambda_i)+\frac{1}{\sqrt{n}}\delta_i\lambda_i-\beta_j)}\\
=&\lambda_i\frac{p-k}{n}\frac{1}{p-k}\sum_{j=1}^{p-k}\frac{\beta_j}{(\psi(\lambda_i)-\beta_j)^2}(1+O_p(\frac{1}{\sqrt{n}}\delta_i))\\
=&\lambda_i\frac{p-k}{n}\int\frac{x}{(\psi(\lambda_i)-x)^2}dF_n(x)(1+O_p(\frac{1}{\sqrt{n}}\delta_i))\\
=&\lambda_i\frac{p-k}{n}\int\frac{x}{(\psi(\lambda_i)-x)^2}dF_c(x)(1+O_p(\frac{1}{\sqrt{n}}\delta_i))(1+O_p(\frac{k}{n}))\\
=&\lambda_i(c+(\frac{p}{n}-c)-\frac{k}{n})m_3(\psi(\lambda_i))(1+O_p(\frac{1}{\sqrt{n}}\delta_i))(1+O_p(\frac{k}{n}))\\
=&\lambda_icm_3(\psi(\lambda_i))(1+O_p(\sqrt{\frac{k}{n}})),
\end{aligned}
\end{equation}
where $m_3(d)=\int\frac{x}{(d-x)^2}dF_c(x)$.

Noting that
\begin{eqnarray*}
\lambda_im_3(\psi(\lambda_i))&=&\lambda_i\int\frac{x}{(\psi(\lambda_i)-x)^2}dF_c(x)\\
&=&\int\frac{\lambda_i}{(\psi(\lambda_i)-x)}\frac{x}{(\psi(\lambda_i)-x)}dF_c(x)\\
&\leq&\frac{\lambda_i}{\psi(\lambda_i)-(1+\sqrt{c})^2}\int\frac{x}{\psi(\lambda_i)-x}dF_c(x)\\
&=&O(1),
\end{eqnarray*}
we have
\[
\frac{1}{n}\tr(\A_n)=O_p(1).
\]

Note that,
\begin{eqnarray*}
&&\frac{1}{n}\tr(\A_n^2)\\
&=&\frac{1}{n}\tr\{\lambda_i\frac{1}{n}\X_2[(\psi(\lambda_i)+\frac{1}{\sqrt{n}}\delta_i\lambda_i)\I-\S_{22}]^{-1}(\psi(\lambda_i)\I-\S_{22})^{-1}\X_2^\top\\
&&\lambda_i\frac{1}{n}\X_2[(\psi(\lambda_i)+\frac{1}{\sqrt{n}}\delta_i\lambda_i)\I-\S_{22}]^{-1}(\psi(\lambda_i)\I-\S_{22})^{-1}\X_2^\top\}\\
&=&\lambda_i^2\frac{1}{n}\tr\{[(\psi(\lambda_i)+\frac{1}{\sqrt{n}}\delta_i\lambda_i)\I-\S_{22}]^{-1}(\psi(\lambda_i)\I-\S_{22})^{-1}\S_{22}\\
&&[(\psi(\lambda_i)+\frac{1}{\sqrt{n}}\delta_i\lambda_i)\I-\S_{22}]^{-1}(\psi(\lambda_i)\I-\S_{22})^{-1}\S_{22}\}\\
&=&\lambda_i^2\frac{1}{n}\sum_{j=1}^{p-k}\frac{\beta_j^2}{(\psi(\lambda_i)+\frac{1}{\sqrt{n}}\delta_i\lambda_i-\beta_j)^2(\psi(\lambda_i)-\beta_j)^2}\\
&=&\lambda_i^2\frac{p-k}{n}\frac{1}{p-k}\sum_{j=1}^{p-k}\frac{\beta_j^2}{(\psi(\lambda_i)+\frac{1}{\sqrt{n}}\delta_i\lambda_i-\beta_j)^2(\psi(\lambda_i)-\beta_j)^2}\\
&=&\lambda_i^2\frac{p-k}{n}\int\frac{x^2}{(\psi(\lambda_i)-x)^4}dF_n(x)(1+O_p(\frac{1}{\sqrt{n}}\delta_i))\\
&=&\lambda_i^2\frac{p-k}{n}\int\frac{x^2}{(\psi(\lambda_i)-x)^4}dF_c(x)(1+O_p(\frac{1}{\sqrt{n}}\delta_i))(1+O_p(\frac{k}{n}))\\
&=&\lambda_i^2(c+(\frac{p}{n}-c)-\frac{k}{n})\int\frac{x^2}{(\psi(\lambda_i)-x)^4}dF_c(x)(1+O_p(\frac{1}{\sqrt{n}}\delta_i))(1+O_p(\frac{k}{n}))\\
&=&\lambda_i^2 cm_4(\psi(\lambda_i))(1+O_p(\sqrt{\frac{k}{n}})),
\end{eqnarray*}
where $m_4(d)=\int\frac{x^2}{(d-x)^4}dF_c(x)$.

Noting that
\begin{eqnarray*}
\lambda_i^2m_4(\psi(\lambda_i))&=&\lambda_i^2\int\frac{x^2}{(\psi(\lambda_i)-x)^4}dF_c(x)\\
&=&\int\frac{\lambda_i^2}{(\psi(\lambda_i)-x)^2}\frac{x^2}{(\psi(\lambda_i)-x)^2}dF_c(x)\\
&\leq&\frac{\lambda_i^2}{(\psi(\lambda_i)-(1+\sqrt{c})^2)^2}\int\frac{x^2}{(\psi(\lambda_i)-x)^2}dF_c(x)\\
&=&O(1),
\end{eqnarray*}
we have
\[
\frac{1}{n}\tr(\A_n^2)=O_p(1).
\]

Hence, we have
\[
\frac{1}{n}\sum_{s=1}^nl_s^2=\frac{1}{n}\tr(\A_n^2)=O_p(1).
\]
where $l_1\geq\cdots\geq l_n$ are the eigenvalues of $\A_n$.

By Lemma \ref{lemma:large deviation2}, we have
\begin{equation}\label{thm3:eq5}
||\frac{1}{n}\bSigma_{11}^{-\frac{1}{2}}\X_1^\top\A_n\X_1\bSigma_{11}^{-\frac{1}{2}}-\frac{1}{n}\tr(\A_n)\I||=O_p(\sqrt{\frac{k}{n}}).
\end{equation}
Combining \eqref{thm3:eq2}, \eqref{thm3:eq3}, \eqref{thm3:eq4}, and \eqref{thm3:eq5}, we have the following spectral decomposition of
\begin{equation}\label{thm3:eq6}
\begin{aligned}
&\bSigma_{11}^{-\frac{1}{2}}[\bm K_n(d_i)-\bm K_n(\psi(\lambda_i))]\bSigma_{11}^{-\frac{1}{2}}\\
=&-\frac{1}{\sqrt{n}}\delta_i\lambda_icm_3(\psi(\lambda_i))(1+O_p(\sqrt{\frac{k}{n}}))\I\\
&-\frac{1}{\sqrt{n}}\delta_i\W(d_i)\diag(w_1(d_i),\ldots,w_k(d_i))\W^\top(d_i),
\end{aligned}
\end{equation}
where $\W(d_i)$ is orthogonal and $\max_{1\leq j\leq k}|w_j(d_i)|=O_p(\sqrt{\frac{k}{n}})$.

Noting that $\frac{p}{n}-c=o(n^{-\frac{1}{2}})$, similar to the proof of \eqref{thm1:eq5} in Theorem \ref{theorem:a.s.consistency}, we have
\begin{equation}\label{thm3:eq7}
\frac{1}{n}\tr(\I+\H_n(d_i))=1+cm_1(\psi(\lambda_i))(1+o(n^{-\frac{1}{2}})).
\end{equation}

Combining \eqref{thm3:eq8}, \eqref{thm3:eq1}, \eqref{thm3:eq6}, and \eqref{thm3:eq7}, we have

\begin{equation}\label{thm3:eq9}
\begin{aligned}
&\bSigma_{11}^{-\frac{1}{2}}(d_i \I-\bm K_n(d_i))\bSigma_{11}^{-\frac{1}{2}}\\
=&\bSigma_{11}^{-1}\psi(\lambda_i)\I+\frac{1}{\sqrt{n}}\delta_i\lambda_i\bSigma_{11}^{-1}-[1+cm_1(\psi(\lambda_i))(1+o(n^{-\frac{1}{2}}))]\I\\
&-\frac{1}{\sqrt{n}}\bSigma_{11}^{-\frac{1}{2}}\bm R_n(\psi(\lambda_i))\bSigma_{11}^{-\frac{1}{2}}+\frac{1}{\sqrt{n}}\delta_i\lambda_icm_3(\psi(\lambda_i))(1+O_p(\sqrt{\frac{k}{n}}))\I\\
&+\frac{1}{\sqrt{n}}\delta_i\W(d_i)\diag(w_1(d_i),\ldots,w_k(d_i))\W^\top(d_i).
\end{aligned}
\end{equation}

Note that
\[
\frac{\psi(\lambda_i)-(1+cm_1(\psi(\lambda_i)))\lambda_j}{\lambda_j}=(\frac{\lambda_i}{\lambda_j}-1)(1+\frac{c}{\lambda_i-1}),
\]
and
\[
\begin{array}{lll}\bSigma_{11}^{-1}=\U\diag(\frac{1}{\lambda_1},\ldots,\frac{1}{\lambda_k})\U^\top.
\end{array}
\]

Multiplying both sides of \eqref{thm3:eq9} by $\U^\top$ from the left and by $\U$ from the right, we have
\begin{eqnarray*}
&&\mid \U^\top\bSigma_{11}^{-\frac{1}{2}}(d_i \I-\bm K_n(d_i))\bSigma_{11}^{-\frac{1}{2}}\U\mid\\
&=&|\diag((\frac{\lambda_i}{\lambda_1}-1)(1+\frac{c}{\lambda_i-1}),\ldots,(\frac{\lambda_i}{\lambda_k}-1)(1+\frac{c}{\lambda_i-1}))\\
                   &&+cm_1(\psi(\lambda_i))o(n^{-\frac{1}{2}})\I+\frac{1}{\sqrt{n}}\delta_i\lambda_i\diag(\frac{1}{\lambda_1},\ldots,\frac{1}{\lambda_k})\\
                   &&-\frac{1}{\sqrt{n}}\U^\top\bSigma_{11}^{-\frac{1}{2}}\bm R_n(\psi(\lambda_i))\bSigma_{11}^{-\frac{1}{2}}\U\\
                   &&+\frac{1}{\sqrt{n}}\delta_i\lambda_icm_3(\psi(\lambda_i))(1+O_p(\sqrt{\frac{k}{n}}))\I\\
                   &&-\frac{1}{\sqrt{n}}\delta_i\U^\top\W(d_i)\diag(w_1(d_i),\ldots,w_k(d_i))\W^\top(d_i)\U|.
\end{eqnarray*}

Note that $\U^\top\W(d_i)$ is an orthogonal matrix and $\max_{1\leq j\leq k}|w_j(d_i)|=O_p(\sqrt{\frac{k}{n}})$.
The $i$-th diagonal element of
\[
\frac{1}{\sqrt{n}}\delta_i\U^\top\W(d_i)\diag(w_1(d_i),\ldots,w_k(d_i))\W^\top(d_i)\U
\]
is $\frac{1}{\sqrt{n}}\delta_iO_p(\sqrt{\frac{k}{n}})$. Since $(\frac{\lambda_i}{\lambda_i}-1)(1+\frac{c}{\lambda_i-1})=0$, by Lemma \ref{lemma:diagonal}, the $i$-th diagonal element of
\[
\A(d_i)=\U^\top\bSigma_{11}^{-\frac{1}{2}}(d_i \I-\bm K_n(d_i))\bSigma_{11}^{-\frac{1}{2}}\U
\]
satisfies
\begin{eqnarray*}
&&\frac{1}{\sqrt{n}}\delta_i(1+cm_3(\psi(\lambda_i))\lambda_i)-[\frac{1}{\sqrt{n}}\U^\top\bSigma_{11}^{-\frac{1}{2}}\bm R_n(\psi(\lambda_i))\bSigma_{11}^{-\frac{1}{2}}\U]_{ii}\\
&&+\frac{1}{\sqrt{n}}\delta_iO_p(\sqrt{\frac{k}{n}})+o(n^{-\frac{1}{2}})=O_p(\sqrt{\frac{k^4}{n^2}}).
\end{eqnarray*}

That is,
\begin{eqnarray*}
&&\frac{1}{\sqrt{n}}\delta_i(1+cm_3(\psi(\lambda_i))\lambda_i)-[\frac{1}{\sqrt{n}}\U^\top\bSigma_{11}^{-\frac{1}{2}}\bm R_n(\psi(\lambda_i))\bSigma_{11}^{-\frac{1}{2}}\U]_{ii}\\
&&+\frac{1}{\sqrt{n}}\delta_iO_p(\sqrt{\frac{k}{n}})+o(n^{-\frac{1}{2}})+O_p(\sqrt{\frac{k^4}{n^2}})=0.
\end{eqnarray*}

Or equivalently,
\begin{eqnarray*}
&&\mid\delta_i(1+cm_3(\psi(\lambda_i))\lambda_i)-[\U^\top\bSigma_{11}^{-\frac{1}{2}}\bm R_n(\psi(\lambda_i))\bSigma_{11}^{-\frac{1}{2}}\U]_{ii}\\
&&+\delta_iO_p(\sqrt{\frac{k}{n}})+o(1)+O_p(\sqrt{\frac{k^4}{n}})\mid=0.
\end{eqnarray*}

Noting that
\[
\delta_i=\frac{\sqrt{n}(d_i-\psi(\lambda_i))}{\lambda_i}=O_p(\sqrt{k}),
\]
under the condition $k=o(n^{\frac{1}{4}})$, $\delta_i$ tends to a solution of
\[
\mid\delta_i(1+cm_3(\psi(\lambda_i))\lambda_i)-[\U^\top\bSigma_{11}^{-\frac{1}{2}}\bm R_n(\psi(\lambda_i))\Sigma_{11}^{-\frac{1}{2}}\U]_{ii}\mid=0.
\]

That is, for any $1\leq i\leq k$,
\[
\delta_i\stackrel{a.s.}{\longrightarrow}(1+cm_3(\psi(\lambda_i))\lambda_i)^{-1}[\U^\top\bSigma_{11}^{-\frac{1}{2}}\bm R_n(\psi(\lambda_i))\bSigma_{11}^{-\frac{1}{2}}\U]_{ii}.
\]
\end{proof}

\begin{proof}[\bf Proof of Corollary \ref{corollary:large deviation1}]
Similar to the proof of Corollary \ref{lemma:large deviation1}, by Chebyshev's inequality, we have
\begin{eqnarray*}
&&P(||\frac{1}{n}\bSigma_{11}^{-\frac{1}{2}}\X_1^\top(\I+\A_n)\X_1\bSigma_{11}^{-\frac{1}{2}}-\frac{1}{n}\tr(\I+\A_n)I||>t)\\
&\leq&P(\max_{1\leq i\leq k}\sum_{1\leq j\leq k}|\frac{1}{n}\sum_{s=1}^n(1+l_s)z_{is}z_{js}-\frac{1}{n}\sum_{s=1}^n(1+l_s)I_{ij}|>t)\\
&\leq&\sum_{1\leq i\leq k}P(\sum_{1\leq j\leq k}\sum_{s=1}^n(1+l_s)|z_{is}z_{js}-I_{ij}|>nt)\\
&=&\sum_{1\leq i,j\leq k}\frac{\sum_{1\leq j\leq k}\sum_{s=1}^n(1+l_s)^2E(z_{is}^2z_{js}^2)}{n^2t^2}\\
&=&\sum_{1\leq i,j\leq k}\frac{\sum_{s=1}^n(1+l_s)^2E(z_{is}^2z_{js}^2)}{n^2t^2}\\
&\leq&\sum_{1\leq i,j\leq k}\frac{\sum_{s=1}^n(1+l_s)^2\sqrt{E(z_{is}^4)E(z_{js}^4)}}{n^2t^2}\\
&\leq&\frac{Ck^2\sum_{s=1}^n(1+l_s)^2}{n^2t^2}.
\end{eqnarray*}

Thus, we have
\[
||\frac{1}{n}\bSigma_{11}^{-\frac{1}{2}}\X_1^\top(\I+\A_n)\X_1\bSigma_{11}^{-\frac{1}{2}}-\frac{1}{n}\tr(\I+\A_n)\I||=\sqrt{\frac{1}{n}\sum_{s=1}^n(1+l_s)^2}O_p(\frac{k}{\sqrt{n}}).
\]
\end{proof}

\begin{proof}[\bf Proof of Theorem \ref{theorem:4}]

Let $\log L_{k'}=-\frac{1}{2}n\sum_{i=1}^{k'}\log d_i-\frac{1}{2}n(p-k')\log\hat{\lambda}_{k'}$. According to \cite{Bai:Choi:Fujikoshi:2018},
\[
\hat{\lambda}=1/(p-k)\sum_{i=k+1}^pd_i\stackrel{a.s.}{\longrightarrow}1.
\]

Suppose $k'<k$. We have
\[
\ell(k)-\ell(k')=\log L_k-\log L_{k'}-\gamma(k-k')(p-(k+k')/2+1/2).
\]

By Theorem \ref{theorem:a.s.consistency}, for $k'<i\leq k$,
\[
\frac{d_{i}}{\psi(\lambda_{i})}\stackrel{a.s.}{\longrightarrow} 1.
\]

Since $\psi(\lambda_k)-1-\log \psi(\lambda_k)>2\gamma c$, by Taylor's expansion, we get
\begin{eqnarray*}
&&P\{\ell(k)>\ell(k')\}\\
&=&P\{\log L_k-\log L_{k'}>\gamma(k-k')(p-(k+k')/2+1/2)\}\\
&=&P\{-\frac{1}{2}n\sum_{i=1}^{k}\log d_i-\frac{1}{2}n(p-k)\log\hat{\lambda}+\frac{1}{2}n\sum_{i=1}^{k'}\log d_i+\frac{1}{2}n(p-k')\log\hat{\lambda}_{k'}\\
&&>\gamma(k-k')(p-(k+k')/2+1/2)\}\\
&=&P\{-\frac{1}{2}n(p-k)\log\hat{\lambda}+\frac{1}{2}n(p-k')\log\hat{\lambda}_{k'}-\frac{1}{2}n\sum_{i=k'+1}^k\log d_i\\
&&>\gamma(k-k')(p-(k+k')/2+1/2)\}\\
&=&P\{-\frac{1}{2}n(p-k')\log\hat{\lambda}+\frac{1}{2}n(p-k')\log\hat{\lambda}_{k'}+\frac{1}{2}n(k-k')\log\hat{\lambda}\\
&&-\frac{1}{2}n\sum_{i=k'+1}^k\log d_i>\gamma(k-k')(p-(k+k')/2+1/2)\}\\
&=&P\{\frac{1}{2}n(p-k')\log(\hat{\lambda}_{k'}/\hat{\lambda})+\frac{1}{2}n(k-k')\log\hat{\lambda}\\
&&-\frac{1}{2}n\sum_{i=k'+1}^k\log d_i>\gamma(k-k')(p-(k+k')/2+1/2)\}\\
&=&P\{\frac{1}{2}n(p-k')\log[(1-(k-k')/(p-k'))(\sum_{i=k'+1}^pd_i/\sum_{i=k+1}^pd_i)]+\frac{1}{2}n(k-k')\log\hat{\lambda}\\
&&-\frac{1}{2}n\sum_{i=k'+1}^k\log d_i>\gamma(k-k')(p-(k+k')/2+1/2)\}\\
&=&P\{-\frac{1}{2}n(k-k')+\frac{1}{2}n(p-k')\log[(\frac{1}{p-k}\sum_{i=k'+1}^kd_i+\frac{1}{p-k}\sum_{i=k+1}^pd_i)/(\frac{1}{p-k}\sum_{i=k+1}^pd_i)]\\
&&+o(n(k-k'))-\frac{1}{2}n\sum_{i=k'+1}^k\log d_i>\gamma(k-k')(p-(k+k')/2+1/2)\}.
\end{eqnarray*}

For the case $\frac{1}{p-k}\sum_{i=k'+1}^k\psi(\lambda_i)=o(1)$, noting that
$\frac{1}{p-k}\sum_{i=k'+1}^kd_i=o_p(1)$, we have
\begin{eqnarray*}
&&P\{\ell(k)>\ell(k')\}\\
&=&P\{-\frac{1}{2}n(k-k')+\frac{1}{2}n(p-k')\log[(\frac{1}{p-k}\sum_{i=k'+1}^kd_i+\frac{1}{p-k}\sum_{i=k+1}^pd_i)/(\frac{1}{p-k}\sum_{i=k+1}^pd_i)]\\
&&+o(n(k-k'))-\frac{1}{2}n\sum_{i=k'+1}^k\log d_i>\gamma(k-k')(p-(k+k')/2+1/2)\}\\
&=&P\{-\frac{1}{2}n(k-k')+\frac{1}{2}n(p-k')[(\frac{1}{p-k}\sum_{i=k'+1}^kd_i)/(\frac{1}{p-k}\sum_{i=k+1}^pd_i)]+o(n(k-k'))\\
&&-\frac{1}{2}n\sum_{i=k'+1}^k\log d_i>\gamma(k-k')(p-(k+k')/2+1/2)\}\\
&=&P\{\frac{1}{2}n\sum_{i=k'+1}^k(d_{i}-1-\log d_{i})+o(n(k-k'))\\
&&>\gamma(k-k')(p-(k+k')/2+1/2)\}\\
&=&P\{\frac{1}{2}n\sum_{i=k'+1}^k(\psi(\lambda_i)-1-\log \psi(\lambda_i))>\gamma(k-k')(p-(k+k')/2+1/2)(1+o(1))\}\\
&\geq&P\{\psi(\lambda_k)-1-\log \psi(\lambda_k)>2\gamma(p-(k+k')/2+1/2)(1+o(1))/n\}\\
&\geq&P\{\psi(\lambda_k)-1-\log \psi(\lambda_k)>2\gamma p(1+o(1))/n\}\\
&\rightarrow&1.
\end{eqnarray*}

For the case $\frac{1}{p-k}\sum_{i=k'+1}^k\psi(\lambda_i)/C \rightarrow 1$, where $C$ is a constant, noting that
$\frac{1}{p-k}\sum_{i=k'+1}^kd_i/C\stackrel{a.s.}{\longrightarrow} 1$, we have
\[
\prod_{i=k'+1}^kd_i\leq (\frac{1}{k-k'}\sum_{i=k'+1}^kd_i)^{k-k'},
\]
\[
\sum_{i=k'+1}^k\log d_i\leq(k-k')\log(\frac{1}{k-k'}\sum_{i=k'+1}^kd_i)=O_p(k\log p),
\]
we have
\begin{eqnarray*}
&&P\{\ell(k)>\ell(k')\}\\
&=&P\{-\frac{1}{2}n(k-k')+\frac{1}{2}n(p-k')\log[(\frac{1}{p-k}\sum_{i=k'+1}^kd_i+\frac{1}{p-k}\sum_{i=k+1}^pd_i)/(\frac{1}{p-k}\sum_{i=k+1}^pd_i)]\\
&&+o(n(k-k'))-\frac{1}{2}n\sum_{i=k'+1}^k\log d_i>\gamma(k-k')(p-(k+k')/2+1/2)\}\\
&\geq&P\{-\frac{1}{2}n(k-k')+\frac{1}{2}n(p-k')\log C+o(n(k-k'))\\
&&-\frac{1}{2}n\sum_{i=k'+1}^k\log d_i>\gamma(k-k')(p-(k+k')/2+1/2)\}\\
&\rightarrow&1.
\end{eqnarray*}

For the case $\frac{1}{p-k}\sum_{i=k'+1}^kd_i/C_n\rightarrow1$, where $C_n\rightarrow\infty$, noting that
$\frac{1}{p-k}\sum_{i=k'+1}^kd_i/C_n\stackrel{a.s.}{\longrightarrow}1$, we have
\[
\sum_{i=k'+1}^k\log d_i\leq(k-k')\log(\frac{1}{k-k'}\sum_{i=k'+1}^kd_i)=O_p(k(\log C_n+\log p))
\]
we have
\begin{eqnarray*}
&&P\{\ell(k)>\ell(k')\}\\
&=&P\{-\frac{1}{2}n(k-k')+\frac{1}{2}n(p-k')\log[(\frac{1}{p-k}\sum_{i=k'+1}^kd_i+\frac{1}{p-k}\sum_{i=k+1}^pd_i)/(\frac{1}{p-k}\sum_{i=k+1}^pd_i)]\\
&&+o(n(k-k'))-\frac{1}{2}n\sum_{i=k'+1}^k\log d_i>\gamma(k-k')(p-(k+k')/2+1/2)\}\\
&\geq&P\{-\frac{1}{2}n(k-k')+\frac{1}{2}n(p-k')\log C_n+o(n(k-k'))\\
&&-\frac{1}{2}n\sum_{i=k'+1}^k\log d_i>\gamma(k-k')(p-(k+k')/2+1/2)\}\\
&\rightarrow&1.
\end{eqnarray*}

Suppose $k'>k$. For $k<i\leq k'=o(p)$,
\[
d_{i}\stackrel{a.s.}{\longrightarrow} \mu_{1}.
\]
According to \cite{Bai:Choi:Fujikoshi:2018}, $\mu_{1}=(1+\sqrt{c})^2$.

Since $\gamma>1/2+\sqrt{1/c}-\log(1+\sqrt{c})/c$, by Taylor's expansion, we get
\begin{eqnarray*}
&&P\{\ell(k)>\ell(k')\}\\
&=&P\{\log L_k-\log L_{k'}>\gamma(k-k')(p-(k+k')/2+1/2)\}\\
&=&P\{-\frac{1}{2}n\sum_{i=1}^{k}\log d_i-\frac{1}{2}n(p-k)\log\hat{\lambda}+\frac{1}{2}n\sum_{i=1}^{k'}\log d_i+\frac{1}{2}n(p-k')\log\hat{\lambda}_{k'}\\
&&>-\gamma(k'-k)(p-(k'+k)/2+1/2)\}\\
&=&P\{-\frac{1}{2}n(p-k)\log\hat{\lambda}+\frac{1}{2}n(p-k')\log\hat{\lambda}_{k'}+\frac{1}{2}n\sum_{i=k+1}^{k'}\log d_i\\
&&>-\gamma(k'-k)(p-(k'+k)/2+1/2)\}\\
&=&P\{-\frac{1}{2}n(p-k')\log\hat{\lambda}+\frac{1}{2}n(p-k')\log\hat{\lambda}_{k'}-\frac{1}{2}n(k'-k)\log\hat{\lambda}\\
&&+\frac{1}{2}n\sum_{i=k+1}^{k'}\log d_i>-\gamma(k'-k)(p-(k'+k)/2+1/2)\}\\
&=&P\{\frac{1}{2}n(p-k')\log(\hat{\lambda}_{k'}/\hat{\lambda})-\frac{1}{2}n(k'-k)\log\hat{\lambda}\\
&&+\frac{1}{2}n\sum_{i=k+1}^{k'}\log d_i>-\gamma(k'-k)(p-(k'+k)/2+1/2)\}\\
&=&P\{\frac{1}{2}n(p-k')\log[(1+(k'-k)/(p-k'))(\sum_{i=k'+1}^pd_i/\sum_{i=k+1}^pd_i)]-\frac{1}{2}n(k'-k)\log\hat{\lambda}\\
&&+\frac{1}{2}n\sum_{i=k+1}^{k'}\log d_i>-\gamma(k'-k)(p-(k'+k)/2+1/2)\}\\
&=&P\{\frac{1}{2}n(k'-k)+\frac{1}{2}n(p-k')\log[(\frac{1}{p-k}\sum_{i=k+1}^pd_i-\frac{1}{p-k}\sum_{i=k+1}^{k'}d_i)/(\frac{1}{p-k}\sum_{i=k+1}^pd_i)]\\
&&-o(n(k'-k))+\frac{1}{2}n\sum_{i=k+1}^{k'}\log d_i>-\gamma(k'-k)(p-(k'+k)/2+1/2)\},
\end{eqnarray*}

\begin{eqnarray*}
&&P\{\ell(k)>\ell(k')\}\\
&=&P\{\frac{1}{2}n(k'-k)-\frac{1}{2}n(p-k')[(\frac{1}{p-k}\sum_{i=k+1}^{k'}d_i)/(\frac{1}{p-k}\sum_{i=k+1}^pd_i)]-o(n(k-k'))\\
&&+\frac{1}{2}n\sum_{i=k+1}^{k'}\log d_i>-\gamma(k'-k)(p-(k'+k)/2+1/2)\}\\
&=&P\{-\frac{1}{2}n\sum_{i=k+1}^{k'}(d_{i}-1-\log d_{i})>-\gamma(k'-k)(p-(k'+k)/2+1/2)(1+o(1))\}\\
&=&P\{-\frac{1}{2}n((1+\sqrt{c})^2-1-2\log(1+\sqrt{c}))>-\gamma(p-(k'+k)/2+1/2)(1+o(1))\}\\
&=&P\{\gamma>\frac{n}{(p-(k'+k)/2+1/2)}(\frac{c}{2}+\sqrt{c}-\log(1+\sqrt{c})(1+o(1))\}\\
&=&P\{\gamma>\frac{p}{(p-(k'+k)/2+1/2)}\frac{n}{p}(\frac{c}{2}+\sqrt{c}-\log(1+\sqrt{c}))(1+o(1))\}\\
&\rightarrow&1.
\end{eqnarray*}
\end{proof}

\bibliographystyle{ims}
\begingroup
\baselineskip=14.5pt
\bibliography{ref}
\endgroup

\end{document}